\def\re#1{(\ref{#1})}
\def\beq{\begin{equation}}
\def\eeq{\end{equation}}
\def\beeq{\begin{eqnarray}}
\def\beeqn{\begin{eqnarray*}}
\def\eeeq{\end{eqnarray}}
\def\eeeqn{\end{eqnarray*}}
\def\nome#1{{\label{#1}}}
\def\ln{large-$N\;$}

\def\de{\delta}                 \def\D{\Delta}

\def\l{\lambda}                 
\def\m{\mu}
\def\n{\nu}

\def\s{\sigma}                  
\def\th{\theta}

\renewcommand{\AA}{{\cal A}}
\newcommand{\DD}{{\cal D}}

\newcommand{\NN}{{\cal N}}
\newcommand{\OO}{{\cal O}}
\newcommand{\PP}{{\cal P}}
\newcommand{\ZZ}{{\cal Z}}
\newcommand{\WW}{{\cal W}}
\newcommand{\Wn}{{\cal W}_n}
\newcommand{\Wh}{\hat{\cal W}_{n}}

\newcommand{\lp}{\left(}
\newcommand{\rp}{\right)}
\renewcommand{\lq}{\left[}
\renewcommand{\rq}{\right]}
\newcommand{\lgr}{\left\{}
\newcommand{\rgr}{\right\}}

\newcommand{\no}{\nonumber}
\newcommand{\ph}{\phantom} 
\def\abs#1{\left|#1\right|}
\def\tr{\,\mbox{Tr}\,}
\def\frac#1#2{ {{#1} \over {#2} }}
\def\fracd#1#2{ {\displaystyle {{#1} \over {#2} }}}
\def\ed#1{\displaystyle{e^{#1}}}
\def\half{\mbox{\small $\frac{1}{2}$}}
\def\p{\partial}

\def\ie{\hbox{\it i.e.}{ }}

\def\bom#1{\mbox{\boldmath$#1$}}
\def\en{\bom{n}}
\def\gq{g^4}

\documentstyle[preprint,aps,amssymb]{revtex}
\begin{document}
\input{epsf}
\draft
\newfont{\form}{cmss10}

\title{Instanton contributions to Wilson loops with general winding number in
two dimensions and  the spectral density}
\author{A. Bassetto$^1$, L. Griguolo$^2$ and
F. Vian$^1$} 
\address{$^1$Dipartimento di Fisica "G. Galilei",
INFN, Sezione di Padova,\\
Via Marzolo 8, 35131 Padua, Italy\\
$^2$Dipartimento di Fisica ``M. Melloni'',
INFN, Gruppo Collegato di Parma, \\ Viale delle Scienze, 43100 Parma, Italy}
\maketitle
\begin{abstract}
The exact expression for Wilson loop averages winding $n$ times on a
closed contour is obtained in two dimensions for pure $U(N)$
Yang-Mills theory and, rather surprisingly, it displays an interesting
duality in the exchange $n \leftrightarrow N$. The large-$N$ limit of
our result is consistent with previous computations. Moreover we
discuss the limit of small loop area ${\cal A }$, keeping $n^2 \AA$ fixed, and
find it coincides with the zero-instanton approximation. We deduce
that small loops, both at finite and infinite ``volume'', are blind to
instantons.
Next we check the non-perturbative result by resumming 't Hooft-CPV and
Wu-Mandelstam-Leibbrandt (WML)-prescribed perturbative series, the former
being consistent with the exact result, the latter reproducing the
zero-instanton contribution. A curious interplay between geometry and
algebraic invariants is observed. 
Finally we compute the spectral density of the Wilson loop operator,
at large $N$, via its Fourier representation, both for 't Hooft and
WML: for small area they exhibit a gap and coincide when the theory
is considered on the sphere $S^2$.

\end{abstract}
\vskip 2.0truecm
DFPD 99/TH 23

\noindent
UPRF-99-08


\vskip 3.0truecm
\vfill\eject

\narrowtext
\section{Introduction}
The non-perturbative structure of non-abelian quantum gauge theories
is still a challenging topic in spite of a large amount of efforts in
this direction. Whereas perturbation theory provides a
well-established computational tool to describe the weak coupling
regime, quantitative predictions for the behaviour of the strongly-coupled 
theory are extremely hard to be found. In the last few years
much attention has been devoted to derive exact results in the
supersymmetric case, by exploiting duality properties
\cite{sei,maldacena}. Nevertheless we believe a ``traditional'' field
theory approach is appealing, most of all in comparing  perturbative
and non-perturbative aspects.

In particular this  approach becomes crucial in the light-front formulation of
quantum gauge theories \cite{light}: though some non-perturbative
features are thought to be transparent, a consistent framework in the
continuum is still lacking. Moreover the relation with the usual
perturbative equal-time quantization remains  unclear. 

Such problems
have recently been tackled in the simplified context of
two-dimensional gauge theories ($YM_2$), taking advantage of the lattice
solutions \cite{migdal}. As far as the continuum is concerned,
in two dimensions the theory seems  trivial when quantized
in the light-cone gauge. As a matter of fact, in the absence of dynamical
fermions, it looks indeed free, being described by a Lagrangian quadratic
in the fields.

Still topological degrees of freedom occur if the theory is put
on a (partially or totally) compact manifold, whereas the simpler 
behaviour enforced by the light-cone gauge condition on the plane
entails a severe worsening in its infrared structure.
These features are related aspects of the same basic issue: even in two
dimensions the theory contains non trivial dynamics, as immediately suggested
by other gauge choices as well as by perturbative calculations of
gauge invariant quantities, typically of Wilson loops\cite{noi1}.
One can say that, in light-cone gauge, dynamics gets hidden
in the very singular nature of correlators at large distances
(IR singularities). 

The first quantity that comes to mind is the 
two-point correlator. If the theory is quantized in the gauge
$A_{-}=0$ at {\em equal-times}, the free propagator has the following
causal expression (WML prescription) in two dimensions
\begin{equation}
\label{WMLprop}
D^{WML}_{++}(x)={1\over {2\pi}}\,\frac{x^{-}}{-x^{+}+i\epsilon x^{-}}\,,
\qquad\qquad x^{\pm}=\frac{x^0\pm x^1}{\sqrt2},
\end{equation}
first proposed by T.T. Wu \cite{wu}. In turn this propagator is nothing
but the restriction in two dimensions of the expression proposed 
by S. Mandelstam
\cite{mandel} and G. Leibbrandt \cite{leib} in four dimensions and 
derived by means of a canonical quantization in ref.\cite{bosco}.
In the equal-time formulation, the causal behaviour is induced by the 
propagation of unphysical degrees of freedom (probability ghosts),
which can  be expunged from the ``physical'' Hilbert space,
but still contribute in intermediate lines 
as timelike
photons do in the QED Gupta-Bleuler quantization.

In dimensions higher than two, where ``physical'' degrees of freedom
are switched on (transverse ``gluons''), this causal prescription is
the only acceptable one; indeed causality is mandatory in order to
get correct analyticity properties, which in turn are the basis of
any consistent renormalization program\cite{libro}.
It has been shown in perturbative calculations \cite{pertu} that agreement 
with Feynman gauge results can only be obtained if a causal propagator is used 
in light-cone gauge.

The situation is somewhat different in exactly two dimensions. Here 
the theory can be quantized on the {\em light-front} (at equal $x^{+}$);
with such a choice, 
no dynamical degrees of freedom occur as the non-vanishing 
component of the vector field does not propagate
\begin{equation}
\label{CPVprop}
D^{P}_{++}(x)=-\frac{i}{2}|x^{-}|\,\delta(x^{+}),
\end{equation}
but rather gives rise to an instantaneous (in $x^{+}$) Coulomb-like
potential. On the other hand, renormalization is no longer a concern 
for a pure Yang-Mills theory in two dimensions.

A formulation based essentially on the potential in Eq.~\re{CPVprop}
was originally proposed by G. 't Hooft in 1974 \cite{hooft}, to derive
beautiful solutions for the $q\bar q$-bound state problem under the form of
rising Regge trajectories.

When inserted in perturbative Wilson loop calculations, expressions 
~\re{WMLprop} and ~\re{CPVprop} lead to completely different
results, as first noticed in ref.\cite{noi1}. The origin of this 
discrepancy was eventually understood in ref.\cite{capo}, where it
was shown that genuine non-perturbative excitations (``instantons'')
are necessary in the equal-time formulation (Eq.~\re{WMLprop})
in order to obtain the  exact result, which in turn is easily
recovered in the light-front formulation (Eq.~\re{CPVprop})
just by summing the perturbative series. This surprising feature 
strengthens the belief 
that the light-front vacuum 
may indeed be much simpler 
than the one in the equal-time formulation, at least in two dimensions. 

In order to gain a deeper insight  in the different physics
described by either  't Hooft or  WML prescription, we have decided to
study a wider class of loops, \ie loops winding around
themselves an arbitrary number of times. Actually, their knowledge
appear to be intimately related\cite{durol} to the general solution of the
Makeenko-Migdal equations at large $N$\cite{migmak}.
Furthermore, in the large-$N$
limit, the spectral density of the eigenvalues of the Wilson operator
is completely determined in terms of such loops. Remarkably this
spectral density carries non-trivial information about the master
field \cite{durol2} of the theory even on the plane.  
Therefore it seems
natural to explore winding loops using both the perturbative WML
prescription and the 't Hooft one in order to understand whether instanton
effects are still able to distinguish  between the two.

In Sect.~II we start by  deriving Wilson loops 
at finite $N$ and arbitrary $n$ on the plane from the non-perturbative
solution on the sphere $S^2$. To our knowledge an explicit  formula, obtained
by decompactification, appears in literature for the first time and
extends previous results limited to a small number of windings
\cite{kazakov,bralic}.   It contains all instanton
contributions and turns out to correspond to the exact resummation of
the perturbative series defined via 't Hooft propagator. Once
again light-front quantization seems to capture the correct vacuum of
the theory. On the other hand, by isolating the zero-instanton
contribution on the sphere, we end up with the perturbative resummation
of WML expressed in terms of a Laguerre polynomial. Hence the picture is
fully consistent with our previous
investigations \cite{capo,grig}.

Having  the general expression for a Wilson loop of $U(N)$ with $n$
windings, we are enabled to perform different limits covering
complementary regimes of the theory. First of all, in the \ln limit
with $g^2N$ fixed, we recover  the well-known result of
\cite{kazakov,rossi} in terms of a Laguerre polynomial. Surprisingly
enough, taking instead $n$
to infinity and keeping $n^2 \AA$ fixed, the perturbative result of WML
is reproduced.
We can explain the latter behaviour by observing  that, having the instantons a
finite size, small loops are essentially blind to them. This  also
teaches us, as we will investigate further  in the sequel, that a
remnant of the weak phase on the sphere survives the decompactification limit.
In fact one can as well study the same limit directly at finite total area
and what is found is precisely the zero-instanton contribution.
The parallelism  of the limits $N\to \infty$ and $n\to \infty$,
reflected by the same functional form, has to be ascribed 
to the symmetry of the general
formula for the Wilson loop under the exchange $N\leftrightarrow n$,
up to a rescaling of the area.

The next step consists in computing the same quantities by means of
the perturbative expansion. This is not simply a check of the
correctness of the previous interpretation, but also 
casts light on the interplay  between geometry and colour in  the 
different  prescriptions, and is done in Sect.~III.
Whereas for WML the resummation of the full perturbative series is a
straightforward generalization of the computation performed in
\cite{tedeschi} for $n=1$, the analog with 't Hooft unveils the full
non-abelian character of such a prescription. Due to its
complexity, we have restricted ourselves to the calculation of the
Wilson loop with winding $n$ to $\OO (g^4)$: as we will see, at this
level already a large amount of classes of diagrams need being
considered. As a matter of fact, we have realized that adopting WML all
the diagrams contributing to a given perturbative order entails the
same geometrical factor, so that only the structure of the colour
traces is relevant. Fortunately the sum of the colour factor is
encoded in the matrix integral of \cite{tedeschi}. At variance with
WML, 't Hooft prescription leads to different geometrical integrals
for different classes of graphs, which, together with the colour
traces, complicate  the resummation of the perturbative series.
Still, in this framework we have been able to describe the mechanism
underlying both the limits  $N\to \infty$ and $n\to \infty$. It is
rather amusing to observe how the latter limit precisely selects a
class of diagram whose geometrical factor is the same as in WML,
explaining in this case the perturbative resummation with WML.
When  performing  the former limit the situation is trivial from an
algebraic point of view (only non-crossed diagrams contribute), while
the  sum over geometrical factors is quite involved. We have evaluated
it by exploiting the
symmetry under the exchange $N\leftrightarrow n$, which shows a
curious interplay between geometry and the algebraic invariants.

Finally in Sect.~IV we have applied previously derived results 
to the computation of the
spectral function in the \ln limit. In the 't Hooft case the
situation is well-known \cite{durol2,gross},  a gap appears
when the area of the loop is below a critical value, resembling
the weak-phase behaviour of the same quantity on the sphere.  The
novelty comes out with WML, for which a gap was expected at any value
of the area, since instanton excitations should be neglected. On the
contrary the gap is present only for sufficiently small loops and, in
this case, again the density coincides with the same quantity on the
sphere in the weak phase.

Conclusions are drawn in Sect.~V, whereas technical details are deferred 
to Appendices A and B.

\section{The Wilson loop with winding number \en}
\noindent
Our starting point are the well-known expressions \cite{migdal} of the
exact partition function and of a  Wilson loop with winding number $n$
for a pure $U(N)$ Yang-Mills theory on a sphere $S^2$ with area $A$
\begin{equation}
\label{partition}
{\cal Z}(A)=\sum_{R} (d_{R})^2 \exp\left[-{{g^2 A}\over 4}C_2(R)\right],
\end{equation}
\beeq
\label{wilson}
&& {\cal W}_n(A-{\cal A},{\cal A})={1\over {\cal Z}N}\sum_{R,S} d_{R} \, d_{S}
\exp\left[-{{g^2 (A-{\cal A})}\over 4}C_2(R)-{{g^2 {\cal A}}\over 4}C_2(S)\right]
\no \\
&& \ph{{\cal W}_n(A-{\cal A},{\cal A})} \times
\int dU {\rm Tr}[U^n]\chi_{R}(U) \chi_{S}^{\dagger}(U),
\eeeq
$d_{R\,(S)}$ being the dimension of the irreducible representation $R(S)$ of
$U(N)$; $C_2(R)$ ($C_2(S)$) is the quadratic Casimir, $A-{\cal A},{\cal A}$
are the
areas singled out by the loop, the integral in (\ref{wilson}) is over the
$U(N)$ group manifold and $\chi_{R(S)}$ is the character of the group
element $U$ in the $R\,(S)$ representation. We immediately notice 
that ${\cal W}_{n}=
{\cal W}_{-n}$.

Eqs.~\re{partition},
\re{wilson} can be easily deduced from the solution of Yang-Mills theory 
on the cylinder, using the fact that the hamiltonian
evolution is governed by the laplacian on $U(N)$: we call Eqs.
(\ref{partition}),(\ref{wilson}) the
heat-kernel representations of ${\cal Z}(A)$ and ${\cal W}_n(A-{\cal A},
{\cal A})$,
respectively.

In order to evaluate ${\cal W}_n(A-{\cal A},{\cal A})$ in the decompactification limit  we
write them explicitly for $N>1$ in the form
\beeq
\label{partip}
&& {\cal Z}(A)=
\frac{1}{N!}\exp \left [ -\frac{g^2 A}{48}N(N^2-1)\right ]
\sum_{m_i = -\infty}^{+\infty}\Delta^2(m_1,...,m_N) \no \\
&&
\times \exp\left [ 
-\frac{g^2A}{4}\sum_{i=1}^N \lp m_i-\frac{N-1}{2} \rp^2\right],
\eeeq
\begin{eqnarray}
\label{wilsonp}
&&{\cal W}_n(A-{\cal A},{\cal A})=\frac{1}{{\cal Z}N N! }\exp \left[-\frac{g^2A}{48}N(N^2-1)
\right ]\\
&\times&\sum_{k= 1}^{N}\sum_{m_i=-\infty}^{+\infty}
\Delta(m_1+n\,\delta_{1,k},...,m_N+n\,\delta_{N,k}) 
\Delta(m_1,...,m_N)\no \\
&\times&
\exp\left [-\frac{g^2 (A-\AA)}{4}\sum_{i=1}^N \lp m_i-\frac{N-1}{2} \rp^2
 -\frac{g^2 {\cal A}}{4}\sum_{i=1}^N \lp m_i-\frac{N-1}{2}+n\,
\delta_{i,k}\rp^2\right].
\nonumber
\end{eqnarray}
We have described the generic irreducible representation by means
of the set of integers $m_{i}=(m_1,...,m_{N})$, related to the
Young tableaux, in terms of which
we get
\begin{eqnarray}
\label{casimiri}
C_2(R)&=&\frac{N}{12}(N^2-1)+\sum_{i=1}^{N}\lp m_{i}-\frac{N-1}{2}\rp^2,
\nonumber \\
d_{R}&=&\Delta(m_1,...,m_{N}).
\end{eqnarray}
$\Delta$ is the Vandermonde determinant and
 the integration in Eq.~(\ref{wilson})
has been performed explicitly, using the well-known formula for the 
characters in terms of the set $m_{i}$.

In this section we will study the partition function \re{partition} and the
Wilson loop \re{wilson} in the limit of infinite area, by showing they
are dominated by
particular representations labelled by suitable indices $\{\hat m_i\}$.
Let us suppose $N$ odd, so that the term $(N-1)/2$ in the Casimir
operator is integer and can be
absorbed in the sum over $m_i$~\footnote{If N is even, $(N-1)/2$ will be
half-integer, but this does not alter our conclusions.}
\beq
C_2(R)=\frac{N}{12}(N^2-1)+\sum_{i=1}^{N} m_{i}^2\,.\nonumber
\eeq
It is now easy to see that the dominant contributions are given by the
following set of indices 
\beq \nome{dom}
\{\hat{m}_i\}=\lgr 0,\, \pm 1, \, \pm 2, \, \cdots,\, \pm \frac{N-1}2 \rgr\,,
\eeq
with all possible permutations,
for which the minimum value of the Casimir is reached and reads
\beq\nome{mincas}
C_2(\{\hat{m}_i\})=\frac{N(N^2-1)}6\,.
\eeq
Finally, by exploiting the symmetry, in the large $A$ limit the
partition function becomes
\beq\nome{partdec} 
\ZZ(A\to \infty ) = \Delta^2(\hat{m}_1,...,\hat{m}_N)
\exp \left[-\frac{g^2A}{24}N(N^2-1)\rq\,.
\eeq
We now turn to the Wilson loop \re{wilsonp}. Thanks to its symmetry,
we can always choose $k=1$ and the equation becomes 
\begin{eqnarray}
\label{wilson1}
&&{\cal W}_n(A-\AA,\AA)=\frac{1}{{\cal Z}}\exp \left[-\frac{g^2A}{48}N(N^2-1)
\right ] \sum_{m_i=-\infty}^{+\infty}
\Delta(m_1,...,m_N)\\
&&\quad \times \Delta(m_1+n,...,m_N)\ 
\exp\left [-\frac{g^2 A}{4}\sum_{i=1}^N {m_i}^2
 -\frac{g^2 \AA}{4} \lp n^2+2n m_1 \rp \right]\,.
\nonumber
\end{eqnarray}
When the previous formula is evaluated in the decompactification limit $A
\to \infty$, $\AA$ fixed, that is
for $\{m_i\}=\{\hat{m}_i\}$, we get
\beq \nome{wildec}
\Wn(\AA;N)=\frac{\exp{\lq -\fracd{g^2 \AA}4 \, n^2 \rq}}{N!} \,
\sum_{\lgr\hat{m}_i\rgr}
\frac{\D(\hat{m}_1+n,...,\hat{m}_N)}{\D(\hat{m}_1,...,\hat{m}_N)}
\exp{\lq -\frac{g^2 \AA }2\, n 
\, \hat{m}_1 \rq} \,.
\eeq
After writing explicitly the Vandermonde determinants and taking
into account the symmetry over the indices
$\{\hat{m}_2,...,\hat{m}_N\}$  we are left with a nice formula for
$\Wn$
\beq \nome{wilnice}  
\Wn(\AA;N)=\frac 1 N \, \exp\lq-\frac{g^2 \AA}4 \, n^2\rq
\sum_{k=-\frac{N-1}2}^{\frac{N-1}2} \exp \lq -\frac{g^2 \AA}2 \,n\,k\rq 
\prod_{{j=-\frac{N-1}2}
\atop{j\neq k}}^{\frac{N-1}2} \frac{k+n-j}{k-j}\,.
\eeq
Remembering that ${\cal W}_{0}=1$ and that ${\cal W}_{n}$ is even in $n$, 
in the following we
shall explicitly consider only positive values of $n$.

It is not difficult to show that \re{wilnice} can be conveniently rewritten 
as
\beeq \nome{wilmorenice1}  
&&\Wn(\AA;N)=\frac{1}{n N} \, \exp\lq- \fracd{g^2 \AA}4 \, 
n(N+n-1)\rq
\sum_{k=0}^{n-1} \fracd{(-1)^k}{k!}
\fracd{(N+n-1-k)!}
{(N-1-k)!(n-1-k)!}\nonumber\\ 
&&\ph{\Wn(\AA;N)=\frac{1}{n N} \, \exp\lq -\fracd{g^2 \AA}4 \, n \rq} \times
\exp \lq \frac{g^2 \AA}2\,n\,k\rq\quad \quad
{\rm for} \quad N>n,
\eeeq
\beeq \nome{wilmorenice2}  
&&\Wn(\AA;N)=\frac{1}{n N} \, \exp\lq-\frac{g^2 \AA}4 \, n(N+n-1)\rq
\sum_{k=0}^{N-1}\fracd{(-1)^k}{k!}\fracd{(N+n-1-k)!}
{(N-1-k)!(n-1-k)!}\nonumber\\ 
&&\ph{\Wn(\AA;N)=\frac{1}{n N} \, \exp[-\fracd{g^2 \AA}4 \, n]} \times
\exp\lq \frac{g^2 \AA}2\,n\,k\rq\quad \quad 
{\rm for} \quad N<n,
\eeeq
that can be combined together to produce the general expression
\beeq \nome{wilmorenice}  
&&\Wn(\AA;N)=\frac{1}{n N} \, \exp\lq-\frac{g^2 \AA}4 \, n(N+n-1)\rq
\sum_{k=0}^{+\infty}\frac{(-1)^k}{k!}\frac{\Gamma(N+n-k)}
{\Gamma(N-k)\Gamma(n-k)}\nonumber\\ 
&&\ph{\Wn(\AA;N)=\frac{1}{n N} \, \exp\lq-\frac{g^2 \AA}4 \, n\rq} \times
\exp\lq\frac{g^2 \AA}2\,n\,k\rq.
\eeeq
The series is actually a finite sum, stopping at $k=n-1$ or $k=N-1$, depending 
on the smaller one. Some comments are now in order. First of all we notice 
that when $n>1$ the simple abelian-like exponentiation is lost. In other 
words the theory starts to feel its non-abelian nature as the appearance 
of different ``string tensions'' makes clear: the Coulomb law is violated 
and the combinatorial coefficients in \re{wilmorenice} are intimately 
related, as we will see in the following, to the presence of instanton 
contributions to light-front vacuum. Actually, from the sphere point of view,
\re{wilmorenice} can be understood as coming from an 
instantons' resummation. Indeed, as first noted by Witten \cite{witte}, it is 
possible to represent ${\cal Z}(A)$ and ${\cal W}_n(A-\AA,\AA)$ as a sum over 
unstable instantons, where each instanton contribution is associated to a 
finite, but not trivial, perturbative expansion. In ref.\cite{capo}, as 
confirmed by the computation on the torus in \cite{grig}, it was shown that 
if only the zero-instanton sector   is considered, in the decompactification 
limit one exactly recovers the sum of the perturbative series for $n=1$, 
in which the light-cone gauge Yang-Mills propagator is WML-prescribed
\re{WMLprop}. Precisely we had for the 
zero-instanton case
\begin{equation}
\label{risultato}
{\cal W}^{(0)}_1=\frac{1}{N}\exp\left[-g^2\frac{(A-\AA)\AA}{4A}\right]\,
L_{N-1}^{(1)}\lp g^2\frac{(A-\AA)\AA}
{2A}\rp.
\end{equation}
$L^{(\alpha)}_\beta(x)$ being the generalized Laguerre polynomials, reproducing 
for $A\to \infty$ the exact resummation of \cite{tedeschi}
\begin{equation}
\label{risultatone}
{\cal W}^{(0)}_1=\frac{1}{N}\exp\left[-\frac{g^2 \AA}4\right]\,
L_{N-1}^{(1)}(g^2 \AA/2).
\end{equation}
This has to be contrasted with the full result coming from \re{wilmorenice} 
\begin{equation}
\label{risultato2}
{\cal W}_1=\exp\left[-\frac{g^2 N \,\AA}4\right]\,,
\end{equation}
giving the expected area-law exponentiation, that can be easily derived 
directly on the plane by resumming the perturbative series in which the 
light-cone propagator is $CPV$-prescribed according 
to 't Hooft \cite{hooft}. Actually, in the large-N limit, \re{risultatone} 
does not exhibit confinement, as first noticed in \cite{tedeschi}. 
As a matter of fact from \re{risultatone}, taking the limit $N\to\infty$ 
($\hat{g}^2=g^2N$),  
\begin{equation}
\label{bessel}
\lim_{N\to \infty} {\cal W}^{(0)}_1=
\sqrt{\frac2{\hat{g}^2 \AA}}
J_1\Bigl(\sqrt{2\hat{g}^2 \AA}\Bigr);
\end{equation}
however, this is hardly surprising since ${\cal W}^{(0)}_1$ does not contain 
any genuine non-perturbative contribution, and simply reproduces the Wilson 
loop behaviour of the weak-coupling phase of the sphere \cite{kaza}, where 
instantons are suppressed \cite{case,gross1,gross}.

\noindent
For a general winding $n$, Eq.~\re{wilmorenice} is therefore expected to come
out from the resummation of 't Hooft perturbative series, that corresponds to a 
light-front quantization of the theory \cite{noi1}. On the other hand to neglect 
instantons, and then to send the area of the sphere to infinity, is likely to 
reproduce the WML computation: the perturbative analysis will confirm these 
claims. At the moment we need the zero instanton contribution to 
\re{wilmorenice}: it can be easily obtained, closely following  ref. 
\cite{capo}, 
by applying a Poisson resummation to \re{wilsonp} and taking the 
non-exponentially suppressed terms as $g^2\to 0$
\begin{equation}
\label{risulta}
{\cal W}^{(0)}_n=\frac{1}{N}\exp\left[-g^2\frac{(A-\AA)\AA}{4A}n^2\right]\,
L_{N-1}^{(1)}\lp g^2\frac{(A-\AA)\AA}{2A}n^2\rp \,.
\end{equation}
For $A\to \infty$ it becomes
\begin{equation}
\label{risultatino}
{\cal W}^{(0)}_n=\frac{1}{N}\exp\left[-\frac{g^2 \AA \,n^2}4\right]\,
L_{N-1}^{(1)}\lp g^2 \AA \,n^2/2 \rp.
\end{equation}
The above result is quite different from \re{wilmorenice} and reflects 
much more dramatically  the same discrepancy found for $n=1$: 
the string tension is independent from $N$ and there is no trace of the
non-abelian  nature, since  different exponential weights do not appear. 
Actually, 
the behaviour of ${\cal W}^{(0)}_n$ (the WML one) is, with respect to the 
fundamental winding $n=1$, ${\it exactly}$ the same as in the abelian $U(1)$ 
theory. It is easily proved for $U(1)$ that
\begin{equation}
{\cal W}_n(\AA;1)=\exp\lq-\frac{g^2\AA\,n^2}{4}\rq,
\end{equation}
the $n$ windings resulting trivially in replacing the charge $g$ by $gn$. 
The WML result satisfies the same rule as it  is clearly seen from 
\re{risultatino}
\beq
\nome{abelia}
{\cal W}^{(0)}_n(g^2\AA;N)={\cal W}^{(0)}_1(g^2n^2\AA;N)\,.
\eeq
In this perspective the exact (\ie full instantons 't  Hooft) result
does not seem related to any simple-minded reduction $U(N) \sim U(1)^N$,
suggested by the abelianization of the theory in axial gauges; rather a
non-trivial (non-abelian) effect is present there. 

We can further observe another basic difference between
Eq.~\re{wilmorenice}   and Eq.~\re{risultatino}: as a matter of fact,
the former shows an interesting symmetry between $N$ and $n$. More
precisely, we have that
\beeq \nome{dual}
 \Wn (\AA;N)&=&\WW_N (\tilde{\AA};n)\\
\tilde{\AA} &=& \frac nN \, \AA \,,\no
\eeeq
a relation that is far from being trivial, involving an unexpected
interplay between the geometrical and the algebraic structure of the
theory.
Looking at Eq.~\re{dual}, the abelian-like exponentiation for $U(N)$
when $n=1$ appears to be connected to the $U(1)$ loop with $N$
windings, the ``genuine'' triviality of Maxwell theory providing the
expected behaviour for the string tension. Moreover we notice the
intriguing feature that the large-$N$ limit (with $n$ fixed) is
equivalent to the limit in which an infinite number of windings is
considered with vanishing  rescaled loop area.
Alternatively, this rescaling could be thought to affect the
coupling constant $g^2 \to \frac{n}{N} g^2$. Eq.~\re{risultatino},
of course, does not exhibit such a fancy behaviour; nevertheless
it is related, in a particular dynamical limit, to the full result, as
we will see in the following. 

Let us discuss in detail the large-$N$ limit of Eq.~\re{wilmorenice}:
it was obtained long ago by Kazakov and Kostov \cite{kazakov}\footnote{The
explicit form for finite $N$ and small $n$ ($n=2,3$) can be deduced
from \cite{kazakov} and \cite{bralic}. The latter used a non-abelian version
of Stokes' theorem, and both are consistent with our general formula
\re{wilmorenice}.}, who solved the Makeenko-Migdal equations in the
case at hand. Coming back to Eq.~\re{wilmorenice1}, we introduce the
function $\Wh(\AA;N)$ from
$$
\Wn(\AA;N) = \exp\lq-\frac{g^2 \AA N n}4\rq\, \Wh(\AA;N) \,,
$$
that can be expressed through a contour integral
\beeq\nome{contour}
&& \Wh(\AA;N) = - \frac1{Nn} \sum_{k=0}^{n-1} \exp \lq  \frac{g^2\,\AA}2 
\,n \lp  k-\frac{n-1}2 \rp \rq \frac{(-1)^k}{k!}\no \\
&&\ph{ \Wh(\AA;N)  } \times
\frac{n!}{(n-1-k)!} \frac1{2\pi i} \oint_C dt\,
t^{k-n} (1-t)^{n-1}
\eeeq
($C$ is a contour surrounding the origin of the complex plane).
The binomial sum can be performed and, after reversing the path around
the pole at $t=1$ and finally changing variable ($z=1+t$), we get
\beq \nome{bella}
\Wh(\AA;N) = \frac{(-1)^n}{2\pi iN} \exp \lq -g^2  \AA\, \frac{n (n-1)}4 \rq
 \oint_C \frac{dz}{(1+z)^N} 
\frac{\lq 1-(1+z) \exp \lp g^2  \AA\, n /2 \rp \rq ^{n-1}}{z^{n+1}}\,.
\eeq
The case $n \geq N$ is easily obtained from Eq.~\re{wilmorenice} by exploiting
the symmetry between $n$ and $N$ (Eq.~\re{dual}).

Eq.~\re{bella} is  a nice representation for the Wilson loop with $n$ windings,
which is particularly suitable to discuss the large-$N$ limit. 
Actually, rescaling $z$ into $\frac zN$, we have
\beq \nome{bellaris}
\Wh(\AA;N) = \frac{(-1)^n}{2\pi i}  
\exp \lq -g^2 \AA\, \frac{n (n-1)}4 \rq
 \oint_C \frac{dz}{(1+\frac zN)^N} 
\frac{\lq N-(N+z) \exp \lp g^2  \AA\, n /2 \rp \rq ^{n-1}}{z^{n+1}}\,,
\eeq
and then, by taking the limit $N \to \infty$, $\hat{g}^2=g^2\, N$
fixed, we arrive at
\beeq\nome{bellafinale}
&&\Wh(\AA;N) =- \frac2{\hat{g}^2 \AA \,n} 
\frac1{2\pi i}  
\oint_C  dx\, \frac{(1+x)^{n-1}}{x^{n+1}} \,
\exp \lq -\frac{\hat{g}^2 \AA\, n\, x}2\rq \no\\
&&\ph{\Wh(\AA;N)}=-\frac2{\hat{g}^2 \AA \,n} \, 
L^{(-1)}_{n-1}\lp \hat{g}^2 \AA \,n /2 \rp\,,
\eeeq
from which we can easily recover the Kazakov-Kostov result
\beq\nome{vecchia}
\Wn(\AA;\infty) = \frac1n 
L^{(1)}_{n-1}\lp \hat{g}^2 \AA\,n /2\rp \,
\exp \lq -\frac{\hat{g}^2 \AA\, n}4 \rq  \,.
\eeq
Using Eq.~\re{dual} we are able to perform another limit, namely $n\to
\infty$ with fixed $n^2\,\AA$  
\beq
\nome{granden}
\lim_{n\to\infty} \Wn(\AA;N)  =
\frac 1N \,
L^{(1)}_{N-1}\lp g^2 \AA\,n^2 /2 \rp \,
\exp \lq -\frac{g^2 \AA\, n^2}4\rq  \,.
\eeq
Eq.~\re{granden} {\em exactly} coincides with the WML result
Eq.~\re{risultatino}: this means that in the small area limit (taking
$n\to \infty$ in order to have a non-vanishing interaction) we
essentially recover the zero-instanton, \ie perturbative,
approximation, even from the exact formula Eq.~\re{wilmorenice}.
The same behaviour will show up when studying the spectral density for
Wilson loop eigenvalues (see Sect.~IV). Once
 again the reason for  the equivalence
of Eqs.~\re{granden} and \re{risultatino} is founded on the
non-trivial geometrical aspects encoded in Eq.~\re{wilmorenice}. In
fact, we recall that in Eq.~\re{risultatino} we have neglected the
instanton contributions: we expect that sufficiently
small loops are unaffected by instantons  due to their having a typical
length scale measured by their size. Eq.~\re{granden} supports this
intuitive argument.

This conclusion is not limited to the decompactified case: we can go
further and prove that, even at finite total area $A$, the exact
heat-kernel expression Eq.~\re{wilsonp} reduces to
Eq.~\re{risulta} when $n \to \infty$~\footnote{We need to have $n
\to \infty$ with $n^2\,\AA $ fixed so that the interaction is
finite.}.
Introducing $\l=n^2\,\AA$, we can write Eq.~\re{wilsonp} as
\beeqn
&&{\cal W}_n {\cal Z} = \exp \left[-\frac{g^2A}{48}N(N^2-1)\rq
\exp\left[-\frac{g^2 \l}{4}\rq
\sum_{m_i=-\infty}^{\infty}  \exp \lq - \frac{g^2\,A}4 \sum_{i=1}^N
m_i^2 -\frac{g^2\,\l}2 \,\frac{m_1}n\rq \no \\
&& \ph{{\cal W}_n {\cal Z}} \times
\Delta^2(m_1,...,m_N) \prod_{j=2}^N \lp 1+ \frac n{m_1-m_j}\rp \,.
\eeeqn
Expanding the product we get
\beeqn
&&{\cal W}_n {\cal Z} = \exp \left[-\frac{g^2A}{48}N(N^2-1)\rq
\exp\left[-\frac{g^2 \l}{4}\rq
\sum_{k=0}^{N-1} \frac{(N-1)!}{(N-1-k)!}\, n^k \\
&&\ph{{\cal W}_n {\cal Z}}\times
\sum_{\hat m_i}
\exp \lq - \frac{g^2\,A}4  \sum_{i=1}^N m_i^2 -\frac{g^2\,\l}2\,\frac{m_1}n\rq 
\Delta^2(m_1,...,m_N) \prod_{j=2}^{k+1} \frac1{m_1-m_j} \no \,.
\eeeqn
We obtain a series in $1/n$ from the expansion of the exponential term
\beeq\nome{ultima}
&&{\cal W}_n {\cal Z} = \exp \left[-\frac{g^2A}{48}N(N^2-1)\rq
\exp\left[-\frac{g^2 \l}{4}\rq
\sum_{l=0}^\infty
\sum_{k=0}^{N-1} \frac{(N-1)!}{(N-1-k)!}\, n^{k-l}
\frac{(-g^2\,\l)^l}{2^l\,l!} \no \\
&&\ph{{\cal W}_n {\cal Z}} \times
\sum_{m_i=-\infty}^{\infty}
\exp \lq - \frac{g^2\,A}4  \sum_{i=1}^N m_i^2 \rq 
\Delta^2(m_1,...,m_N) \prod_{j=2}^{k+1} \frac{m_1^l}{m_1-m_j} \,.
\eeeq
Eq.~\re{ultima} is truly an expansion in $1/n^2$: in order to
understand this point, we simply notice that changing $m_i \to - m_i$
we produce an overall factor $(-1)^{k+l}$ weighting the sum over
$m_i$. This implies that $k+l$ (and therefore $k-l$) must be an even
integer so as to have a non-vanishing result. It is useful to rewrite
the sum over $m_i$ as 
$$
\frac1{(k+1)!}
\sum_{m_i=-\infty}^{\infty}
\exp \lq - \frac{g^2\,A}4  \sum_{i=1}^N m_i^2 \rq 
\Delta^2(m_1,...,m_N) \, \sum _{P} \prod_{j=2}^{k+1} 
\frac{m_{P(1)}^l}{m_{P(1)}-m_{P(j)}} \,,
$$
where the sum over $P$ is over the elements of the permutation group
$S_{k+1}$.
Let us evaluate to the $0^{\rm th}$-order term: we observe that for
$l=k$
\beq\nome{sumprod}
\sum _{P} \prod_{j=2}^{k+1} 
\frac{m_{P(1)}^l}{m_{P(1)}-m_{P(j)}} = \frac1{\D(m_1,...,m_{k+1})}\,
\sum_P f_P(m_1,...,m_{k+1})\,,
\eeq
the Vandermonde determinant being produced by the common denominator;
the quantity
$\sum_P f_P(m_1,...,m_{k+1})$ is a polynomial of degree
$\frac{k(k+1)}2$ in $k+1$ variables. Since the Vandermonde determinant 
$\D(m_1,...,m_{k+1})$ is totally antisymmetric, the non-vanishing
contribution to Eq.~\re{ultima} comes from the totally antisymmetric
part of $\sum_P f_P(m_1,...,m_{k+1})$, implying that
$$
\sum_P f_P(m_1,...,m_{k+1}) \simeq c \, \D(m_1,...,m_{k+1})\,.
$$
The constant $c$ is easily proven to be 1 by inspection    of Eq.~\re{sumprod}.
Hence the   contribution of order $\lp\frac1{n^2}\rp^0$ to
Eq.~\re{ultima} is 
\beeq
&&{\cal W}_n {\cal Z} =  
\exp \lq-\frac{g^2A}{48}N(N^2-1)\rq
\lq\sum_{k=0}^{N-1} \frac{(N-1)!}{(N-1-k)!} \, \frac{(-1)^k}{k!\,
(k+1)!}\, \lp \frac{g^2\, \l}2 \rp^k\rq \no \\
&& \ph{{\cal W}_n {\cal Z} } \times \sum_{m_i=-\infty}^\infty
\exp \lq - \frac{g^2\,A}4  \sum_{i=1}^N m_i^2 \rq 
\Delta^2(m_1,...,m_N) \no \\
&& \ph{{\cal W}_n {\cal Z} } =
{\cal Z}[A] \,
\frac1N \, L^{(1)}_{N-1}(g^2\,\l/2)  \, e^{-\frac{g^2\,\l}4} \,,
\eeeq
proving that the $0^{\rm th}$-order contribution is exactly the WML
one.
It remains to show that the divergent contributions (coming from $l<k$ in
Eq.~\re{ultima}) are vanishing. To this purpose we use again
Eq.~\re{sumprod} to notice that we can still factorize a Vandermonde
determinant in the denominator, but now $\sum_P f_P (m_1,...,m_{k+1})$
is a polynomial of degree less than $\frac{k(k+1)}2$ in $k+1$
variables. Its totally antisymmetric part vanishes, so that
$$
\lim_{{n \to \infty}\atop { \l \; {\rm fixed}}}
\Wn (A-\AA,\AA) \sim
\frac1N
 \, L^{(1)}_{N-1}(g^2\,\l/2)\, \exp \lq -\frac{g^2\,\l}4 \rq + \OO
\lp \frac1{n^2} \rp \,.
$$
This confirms our guess that sufficiently small loops do not see
instantons: actually it is known that in the strong phase (at large-$N$)
small loops behave as in the weak phase (\ie they do not confine) as
$n \to \infty$. Our result goes further, revealing that even at finite
$N$ and at finite total area $A$, instantons can be dynamically suppressed. 

\section{The perturbative WML and 't Hooft prescriptions}
\noindent
We now turn to the resummation of the perturbative expansion for
$\Wn$. The aim of our investigation is not only to check the
correctness of the interpretation of the non-perturbative results
derived so far, but more importantly to observe and explain a curious
interplay between geometry and colour factors in the two different
prescriptions. 
We will begin with WML, for which the resummation of the perturbative
series is found through a simple generalization of the  procedure
outlined in \cite{tedeschi} in the case $n=1$. This is due to all diagrams
entailing the same geometrical factor. 
We will then switch to 't Hooft prescription and realize the
resummation is rather involved since  graphs  belonging to the same
class, according to their colour factor, produce different geometrical
integrals. Hence, in this respect, light-front quantization exhibits its
peculiar non-abelian character.

\subsection{Resummation of the perturbative series defined via WML
prescription}

We closely follow the procedure outlined in \cite{tedeschi}. In the
euclidean space the perturbative expansion of $\Wn$ is
\beq
\Wn[\AA]
=1 + \frac1N \sum_{k=1}^\infty (-g^2)^k \int_0^1 ds_1 \;
\dot{x}^{\m_1}(s_1)
\cdots  
\int_0^{s_{2k-1}} ds_{2k} \; \dot{x}^{\m_{2k}}(s_{2k}) \, \tr \lq
G_{\m_1\cdots\m_{2k}} (x_1, \cdots,x_{2k}) \rq\,,
\eeq
where $x^\m (s)$, $s \in [0,1]$ parametrizes the contour with $n$
windings.
The Lie algebra-valued $2k$-point Green function $G_{\m_1\cdots\m_{2k}} 
(x_1, \cdots,x_{2k})$ has to be expressed via the Wick rule in terms of
the free propagator $D(x_i-x_j)$, in the light cone gauge $A_-=0$ and
endowed  with (Wick-rotated) WML prescription 
\beq\nome{propWML}
D_{++}^{ab}(x-x')=\frac1{2\pi} \, \de^{ab} \, \frac{x_+ - x'_+}{x_- -
x'_-}\,.
\eeq
The simplest choice for the contour is a circle wrapping around
itself $n$ times. Then it happens \cite{tedeschi} that the weighted basic 
correlator is independent of the loop variables, since
$$
\dot{x}_-(s) \,\dot{x}_-(s') \, \frac{x_+ (s)- x_+ (s')}{x_- (s) -
x'_-(s')}=2(n\pi r)^2\,.
$$
After that, the integration over the path parameters $s_1,\cdots,s_{2k}$
trivially  yields $1/(2k)!$ and  the task of determining the
Wilson loop reduces to the purely combinatorial problem   of finding
the group factors corresponding to the Wick contractions. 
What is left is
$$
\Wn[\AA]
=1+\frac1N \sum_{k=1}^\infty \lp - \frac{ g^2\, n^2\,\AA}{2}\rp^k
\,\frac{c_{2k}(N)}{(2k)!}\,\,,
$$
where  $c_{2k}(N)$ is the sum over all possible traces of
$2k$ $T^a$ matrices suitably contracted. Fortunately this group factor
is generated by a matrix integral \cite{tedeschi}, and the final result reads
\beq\nome{finalML}
\Wn[\AA]=
\frac1N \exp \lq -\frac14 g^2 n^2 \AA \rq \, L^{(1)}_{N-1} (g^2 n^2 \AA/2)\,,
\eeq
reproducing the zero-instanton result Eq.~\re{risultatino} obtained via
non-perturbative methods.

\subsection{Resummation of the perturbative series defined via 't
Hooft prescription}
\noindent
As already hinted, we were not able to resum the full perturbative
expansion for $\Wn$ prescribed with 't Hooft. We believe after
exertion that such a formidable task is, if not impossible, at least
extremely tough. Nevertheless we were successful  in  the
computation of the Wilson loop with $n$ windings at $\OO (g^4)$, which
suffices to give a flavour on how things work at higher
orders. Moreover we  performed the limit $n \to \infty$, $n^2\AA$
fixed, and, by exploiting the duality $n \leftrightarrow N$, the
large-N limit, with $g^2 N$ fixed.

Firstly, let us start from the perturbative definition  of $\Wn$ in
the light-cone gauge ($A_-=0$)
\beeq\nome{wper}
&&\Wn= \frac1N \NN \tr \lgr
\int \DD A_\m \, \exp\lq i \int d^2x \, \lp -\frac14
F_{\m\n}^a \, F^{\m\n\,a}+J_\m^a\, A^{\m\,a}\rp \rq \right. \no \\
&& \left. \ph{\Wn=\NN \int \DD A_\m \,} \de(A_-^a) \; {\cal P} \exp \lp i\,g
\oint_n A_\m^a \, T^a \, dx^\m \rp \rgr \,,
\eeeq
evaluated at $J^a=0$.
It easy to recognize that Eq.~\re{wper} can be rewritten as follows
\beq\nome{wcurr}
\Wn=\frac1N \,\NN \tr \lgr 
\, {\cal P} \exp \lq g \oint_n T^a \frac{\de}{\de\, J^a(x)} \,
dx^{+}\rq 
\exp \lq -\half \int d^2x \, d^2y \,   J^a(x) \, D(x-y)\,
J^a(y) \rq  \rgr_{J=0} \,,
\eeq
where now the propagator $D(x-y)$ is defined through the CPV prescription
Eq.~\re{CPVprop}.
We consider a light-like rectangle with sides $2L$, $2T$ (see Fig.~1)
and choose the currents with support on the contour, so that
$$
J^a(x^+,x^-)= j_u^a \, \de(x^- - L) + j_d^a \, \de(x^- + L)\,.
$$
With this choice the perturbative expansion Eq.~\re{wcurr} for $\Wn$
reads
\beeq\nome{variexp}
&&\Wn= \frac1N \, \NN \tr \Biggl\{
{\cal P} \exp \lq g \oint_{C_{\,2,n}} T^{a_n} j_d^c
\frac{\de}{\de\, j_u^{a_n}(x^+)} \, dx^{+}\rq 
{\cal P} \exp \lq g \oint_{C_{\,1,n}} T^{b_n} \frac{\de}{\de\,
j_d^{b_n}(x^+)}\, dx^{+}\rq  \no \\
&&\ph{\Wn=\frac1N \, \NN \quad}\cdots {\cal P} \exp \lq g \oint_{C_{\,2,1}}
T^{a_1} 
\frac{\de}{\de\, j_u^{a_1}(x^+) }\, dx^{+}\rq 
{\cal P} \exp \lq g \oint_{C_{\,1,1}} T^{b_1} 
\frac{\de}{\de\, j_d^{b_1}(x^+)}\, dx^{+}\rq \no \\
&& \ph{\Wn=\frac1N \, \NN \cdots \qquad} \exp \lq iL\int^T_{-T}
dx^+\, j_u^c(x^+)\, 
j_d^c(x^+) \rq \Biggr\}_{j=0}\, 
\eeeq
with $C_{\,1}$ and $C_{\,2}$ as in Fig.~1.
Clearly in the limit $j_u^a$, $j_d^a \to 0$  the only non-vanishing
contributions are those with a matching number of derivatives with
respect to $j_u$ and $j_d$.

Now everything is settled and we can easily derive the expression of
$\Wn^{(4)}$, \ie the Wilson loop with $n$ windings $\OO (g^4)$. The
following prefactor is common to all classes of diagrams
\beq\nome{prefactor}
g^4\,(i\,L)^2\,(2\,T)^2=-\frac{g^4\,\AA^2}{4}\,,
\eeq
$\AA=4LT$ being the area of the loop. 
As a matter of fact, $g^4$ comes from four derivatives with respect to the
currents which contribute at this order, $(i\,L)^2$ is produced when two
derivatives act on the last exponential in Eq.~\re{variexp}, which
represents the only non-vanishing contribution, and finally $(2\,T)^2$
is given by integration over the loop variables.
Notice the expected dependence on the area due to invariance with
respect to area-preserving diffeomorphisms typical of $d=2$.

In Appendix A we show which classes of non-crossed and crossed
diagrams contribute to $\Wn^{(4)}$ and also provide the exact counting
(as a function of $n$) and area factor. The latter turns out to be
either $\half$ or $1$ depending on the presence of integrals in the loop
variables which are nested as a consequence of the
definition of the $\PP$ exponential.
Furthermore one has to take into account that non-crossed diagrams
produce an additional factor $N^3/4$ from $\tr \lq T^a\,T^a\,T^b\,T^b
\rq$, whereas the analog for crossed diagrams is $N/4$ from $\tr \lq
T^a\,T^b\,T^a\,T^b\rq$. Remember eventually the factor $1/N$ appearing in
Eq.~\re{variexp}.

We present here the final results, which read
\beeq\nome{resu4}
&&\Wn^{(4),\,nc}=-\frac{g^4\,\AA^2\,n^2}{96} N^2 \lp 2n^2+1\rp\\
&&\Wn^{(4),\,c}=-\frac{g^4\,\AA^2\,n^2}{96} \lp n^2-1\rp
\eeeq
and, summing up,
\beq\nome{sum4}
\Wn^{(4)}=-\frac{g^4\,\AA^2\,n^2}{96} \lp 2n^2 N^2 +N^2+n^2-1 \rp\,,
\eeq
which coincides with the coefficient  $\OO (g^4)$ of the expansion of
Eq.~\re{wilmorenice} obtained via non-perturbative methods. Moreover,
already at this
level, the symmetry under the exchange $n \leftrightarrow N$ is
manifest once the area is rescaled by  $N/n$.

Let us now go back to Eq.~\re{variexp} in order to extrapolate the
relevant limits $n \to \infty$ and $N \to \infty$. We have
already stressed that they are dual to each other (provided the loop
area is suitably rescaled), so that we will explicitly perform  only
the former and deduce the latter by symmetry.
In Sect.~II the large-$n$ limit of the non-perturbative $\Wn$, 
with $\AA n^2$ fixed, was proved to coincide with the resummation of the
perturbative series when WML prescription was adopted and $\AA\to \AA
n^2$. Thus we guess that at large $n$ dominant configurations of loop
diagrams where 't Hooft prescription is imposed must be weighted by
the same area factor. This precisely  happens for WML as we realized
in Sect.~IIIA. In fact, by inspecting the computation of $\Wn$ at
order $\OO (g^4)$ we are easily convinced that in this limit dominant
configurations, either crossed or non-crossed, are those which spread 
through the maximum number of
sheets of the n-fold loop, \ie those in which the vector
fields are all attached on different sides of the loop. For instance
the leading terms at $\OO (g^4)$ are those shown in Figs.~1i,
2d. Going further, at a generic order $g^{2k}$ one needs at least $2k$ sheets 
in order to set up such configurations, and their number is
straightforwardly determined. Actually, there are 
${n \choose 2k}$
ways of extracting $2k$ sheets out of  $n$ and in addition we have to
take into account all possible configurations of the basic $2k$-sheet
structure with any kind of cross-over. Therefore, the counting of
dominant configurations is given by
\beq \nome{coeffi}
\frac{n!}{(2k)!\,(n-2k)!} \times  c_{2k}(N)\,,
\eeq
where $c_{2k}(N)$ is the coefficient we acquainted with in
Sect.~IIIA. Hence, at the leading order in $n$, we end up with
\begin{equation}
\label{leado}
W_{n\to \infty} (\AA)=\frac1N \sum_{k=0}^{\infty} \lp-\frac{g^2 \, \AA\, n^2}
{2}\rp^k \,
\frac{c_{2k}(N)}{(2k)!}
\eeq
and, switching to the euclidean space via a Wick rotation
\begin{equation}
\label{Wick}
W_{n\to \infty} (\AA)=\frac1N \exp \lq -\frac14 \,g^2 \, \AA\, n^2 \rq
\, L^{(1)}_{N-1} \lp \frac{g^2 \, \AA\, n^2}2 \rp \,,
\eeq 
which corresponds to Eq.~\re{risultatino}.

The next step consists in performing the large-$N$ limit of the
resummed perturbative  series in the axial gauge with the light-cone
CPV prescription. The duality $n \leftrightarrow N$ allows us to conclude
\begin{equation}
 \label{conclu}
W_{N\to \infty} (\AA)=\frac1n \exp \lq -\frac14 \,\hat g^2 \, \AA\, n \rq
\, L^{(1)}_{n-1} \lp \frac{\hat g^2 \, \AA\, n}2 \rp \,,
\eeq
where the rescaling in Eq.~\re{dual} is understood.
Apart from the numerical coincidence, the duality entails a deep
relation among different faces of the same theory. Naively one could think
that the large-$N$ limit, being dominated by non-crossed graphs
in which the trace factor considerably simplifies, is
easy to recover . Nevertheless,  a surprising complexity in the integrals
over loop variables arises, since every class of diagrams is
characterized by a different area factor. This is already manifest from the
computation of the Wilson loop  at order $g^4$ (see Appendix A).
Thus the most relevant conclusion we can draw is that an enlightening
interplay between geometry and algebraic invariants  takes place,
connecting the large-$N$ with the large-$n$ limit and, in a sense, the
strong and the weak phase of the theory.

\section{The spectral density}
\noindent
$U(N)$ Yang-Mills theory on the sphere $S^2$ exhibits in the large-$N$
limit with $g^2\,N$ fixed, a third order phase transition 
(Douglas-Kazakov (DK) phase transition)\cite{kaza}. According to the total area
$A$ of the sphere, two different regimes occur: a strong coupling 
regime when $A>\pi^2$ (in suitable units, $i.e.$ $\frac{g^2\,N}{2}=1$) and a 
weak coupling regime with $A<\pi^2$. The difference between them 
can be traced back to the presence of instantons on the sphere
\cite{case,gross1,gross}, which fully contribute for large $A$, but are instead
exponentially suppressed in the weak-coupling phase.

The two regimes can be conveniently characterized by the density $\rho$
of Young tableaux indices which single out the various irreducible 
representations of the group $U(N)$. In the strong coupling regime
the distribution $\rho$ undergoes 
a phenomenon of ``saturation'' in the large-$N$ limit,
namely it is equal to one in a finite interval of its domain. 
In turn, if we consider a Wilson 
loop on the sphere with the contour $C$ running around an
equator and winding an arbitrary number $n$ of times,
the density $\rho$ is related by a kind of duality transformation\cite{gross}
to the spectral density 
$\sigma$ on the unit circle of the eigenvalues 
of the unitary operator $U_{C}$, representing the loop.

The strong-coupling phase is characterized by the fact that those
eigenvalues completely fill the unit circle, whereas in the weak coupling
regime a gap develops where the distribution $\sigma$ vanishes.
Before decompactifying the sphere to the plane by taking the limit
$A\to \infty$, we have to learn how the density $\sigma$ evolves
when we replace the equatorial contour with a smooth arbitrary one.
In ref. \cite{gross} it is shown that, when one of the two different areas 
generated in this way, 
let us
call it ${\cal A}$, becomes
small enough, a gap in the eigenvalue distribution appears even
in the strong DK phase ($i.e.$ for $A>\pi^2$). Actually, when the total area
$A$ approaches its critical value $\pi^2$ from above, ${\cal A}$
becomes critical at its maximum value ${\cal A}_{cr}=\frac {\pi^2}{2}$,
namely when the contour is along an equator. 

When the sphere is decompactified to the plane by taking the limit
$A\to \infty$ at fixed ${\cal A}$, we are automatically in the DK strong 
phase; it is then remarkable that ${\cal A}_{cr}$
reaches, in the limit
$A\to \infty$, a well defined value, ${\cal A}_{cr}=4$.

We are going to obtain all these results from the explicit expressions
we got in Sect.~II for Wilson loops with an arbitrary winding number 
${\cal W}_{n}$.
We shall explicitly discuss the large-$N$ case on the plane ($A\to \infty$).
The spectral density $\sigma({\cal A},\theta)$, ${\cal A}$ being the area
enclosed by the loop, is related to ${\cal W}_{n}$ by the following Fourier 
series
\cite{durol}
\begin{equation}
\label{fourier}
\sigma({\cal A},\theta)=\frac{1}{2\pi}\,\lq 1+2\sum_{n=1}^{\infty}
\cos(n\theta)\,{\cal W}_{n} \rq \,.
\end{equation}
We shall compute this function and show that, according to different values
of ${\cal A}$, a gap may develop in the $\theta$-variable in which the
distribution vanishes. This phenomenon was first noticed by Durhuus and
Olesen\cite{durol2}, who derived the spectral function solving a kind of
Makeenko-Migdal equation with suitable boundary conditions.

By introducing Eq.~\re{vecchia} into Eq.~\re{fourier} we get
\begin{equation}
\label{laguerre}
\sigma({\cal A},\theta)=\frac{1}{2\pi}\,\lq 1+\sum_{n=1}^{\infty}
(\ed{in\theta}
+\ed{-in\theta})\frac{1}{n}\,\ed{-\frac{n{\cal A}}{2}}\,L_{n-1}^{(1)}
(n{\cal A})\rq
=\frac{1}{2\pi} \lq 1+2 Re\,F({\cal A},\theta)\rq \,,
\end{equation}
where
\begin{equation}
\label{rappres}
F({\cal A},\theta)=\sum_{n=1}^{\infty}
\frac{1}{n}e^{-n(\frac{{\cal A}}{2}-i\theta)}\oint_{\Gamma}
\frac{dt}{2\pi i}\,e^{-n{\cal A} t}\lp 1+\frac{1}{t}\rp^n\,.
\end{equation}
Here we have used a well-known integral representation for the Laguerre
polynomials in which the integration contour $\Gamma$
encircles the origin of the complex $t$-plane.

When the inequality
\begin{equation}
\label{ineq}
\abs{e^{-{\cal A}\,(t+\frac{1}{2})}\,\lp 1+\frac{1}{t}\rp}<1
\end{equation}
is satisfied, the series in Eq.~\re{rappres} can be summed giving rise to a
{\em single-valued} function
\begin{equation}
\label{single}
\sum_{n=1}^{\infty}
\frac{1}{n}e^{-n\lq{\cal A}(t+\frac{1}{2})
-i\theta\rq}\,\,\lp 1+\frac{1}{t}\rp^n=-\log\,\lq 1-
e^{-{\cal A}\lp t+\frac{1}{2}
\rp +i\theta}\,\,\lp 1+\frac{1}{t}\rp\rq.
\end{equation}
The logarithm in turn can be analytically continued; as is well known, it has branch points when its argument either vanishes or diverges. Whereas it is easy
to realize that the only divergence occurs, at finite $|t|$, at $t=0$,
a careful study of the roots of the trascendental equation
\begin{equation}
\label{trasc}
\ed{-{\cal A} (t+\frac{1}{2})
+i\theta}\,\lp 1+\frac{1}{t}\rp=1
\end{equation}
is in order before drawing any conclusion.

It is amusing to notice that the change of variable
\begin{equation}
\label{change}
t=\frac{\exp(i\xi)}{1-\exp(i\xi)}
\end{equation}
turns Eq.~\re{trasc} into Eq.(3.4) of ref.\cite{durol2}.

To the success of our calculation it is essential that the contour $\Gamma$
in Eq.~\re{rappres} can be deformed while the logarithm still remaining
single-valued on it.

The discussion of the roots of Eq.~\re{trasc} is deferred to Appendix B.
There we show that, for ${\cal A}>4$, only two roots are possible and
only one of them is encircled by the contour $\Gamma$. If we set $t=x+iy$ 
and denote this root by the values $(\hat x, \hat y)$, which obviously are
 functions of ${\cal A}$ and of $\theta$, a cut can be drawn along the segment 
$(0,0)$---$(\hat x,\hat y)$ in the 
complex $t$-plane and the contour $\Gamma$
can be deformed to the contour $\gamma$, 
encircling the branch points and just running 
below and above the cut. It is then immediate to conclude
that $$F({\cal A},\theta)=\hat x+i\hat y$$ and that, eventually,
\begin{equation}
\label{sol}
\sigma({\cal A},\theta)=\frac{1}{\pi}\lp\hat x +\frac{1}{2}\rp.
\end{equation}

As long as ${\cal A}>4$, $\hat x+\frac{1}{2}$ is positive  
for any value of
$\theta$. The eigenvalues are filling the entire interval $(-\pi,\pi)$.

When ${\cal A}<4$ the situation dramatically changes. 
Although we are in the
strong DK phase (we are on the plane), a gap starts showing up in the
eigenvalue density for $\theta$ close to $\pi$ ( and to $-\pi$; remember
$\sigma$ is an even function of $\theta$). Actually, for ${\cal A}<4$,
there is a value $\theta_{cr}<\pi$ such that $\hat x({\cal A},\theta_{cr})
=-\frac{1}{2}.$ Once this value of $\theta$ is reached,
$\hat x$ remains equal to $-\frac{1}{2}$, and consequently $\sigma$ vanishes.

In Appendix B we show that the expression for $\theta_{cr}$ coincides with 
the one in Eq.(4.8) of ref.\cite{durol2}, namely
\begin{equation}
\label{gap}
\theta_{cr}=\sqrt{{\cal A}-\frac{{\cal A}^2}{4}}+\arccos\lp 
1-\frac{{\cal A}}{2}\rp.
\end{equation}
Eventually, in the limit ${\cal A}\to 0$, the density approaches the periodic 
$\delta$-distribution, as it should.

Actually we have already seen in Sect.~II that, in the small-${\cal A}$ limit,
instantons on the sphere are suppressed even in the strong DK phase.
This phenomenon has a simple geometrical interpretation: it occurs when one of
the two areas singled out by the loop
gets small when compared to the instanton size. It might then seem 
that the different behaviour of $\sigma$ according to the values of ${\cal A}$,
is fully driven and characterized by instantons.
This issue would in turn be of crucial
relevance in singling out peculiar properties of the light-front
vacuum when compared to the one in equal-time canonical Fock space
after decompactification of the sphere to the plane. 

Unfortunately the situation is not as simple as that. In ref.\cite{capo} the 
contribution to the Wilson loop from the zero-instanton sector on $S^2$
was clearly pointed out. It was also noticed that, in the decompactification 
limit, this contribution exactly coincides with the sum of the perturbative 
series when graphs are evaluated according to the WML prescription for the
propagator. In Sect.~II we have generalized this result  to a loop with an 
arbitrary winding number $n$. It is then a fairly simple exercise to take 
the large-$N$ limit and to insert the result in Eq.~\re{fourier}
\begin{equation}
\label{zerosec}
\sigma^{\tiny{ML}}=\frac{1}{2\pi}\, \lq1+2\,
\sum_{n=1}^{\infty}\frac{cos(n\theta)}
{n\sqrt{{\cal A}}}\,J_1(2n\sqrt{{\cal A}})\rq, 
\qquad\quad -\pi\le \theta\le \pi,
\end{equation}
$J_1$ being the usual Bessel function. Of course its geometrical meaning 
is now completely different.

Still, due to the absence of instantons, 
one would naively believe that $\sigma^{\tiny{ML}}$
should behave like in a weak phase, in particular should develop a gap.
But this is not always the case; as a matter of fact, using a standard integral
representation for the Bessel function, we get
\begin{equation}
\label{sigmaML}
\sigma^{\tiny{ML}}=\frac{2}{\pi}\int_{-1}^{1} dt \, \sqrt{1-t^2}\, \delta_P(\theta+
2t\sqrt{{\cal A}}),
\end{equation}
$\delta_P$ being the periodic $\delta$-distribution, leading to
\begin{equation}
\label{sigmasomML}
\sigma^{\tiny{ML}}=\frac{1}{\pi\sqrt{{\cal A}}}\sum_m 
\sqrt{1-\frac{(2\pi m+\theta)^2}{4{\cal A}}}\,,
\qquad\quad -\pi\le \theta\le \pi,
\end{equation}
the sum running over all positive and negative values of $m$ such that
the argument of the square root is positive.

It is easy to check that this expression reproduces the 
limits $$\lim_{{\cal A}\to \infty}\sigma^{\tiny{ML}}=\frac{1}{2\pi}$$
and $$\lim_{{\cal A}\to 0}\sigma^{\tiny{ML}}=\delta(\theta).$$

For ${\cal A}<\frac{\pi^2}{4}$ there is indeed a gap, similar to the
one in the weak phase on the sphere, but with a new
threshold. For larger 
values the gap disappears.

Only the functional form (a square root)
is reminiscent of a weak phase. There are moreover discontinuities in
the derivative of the density anytime a new value of $m$ starts contributing.

One would eventually conclude that the occurrence of a gap is not
directly related to the presence of instantons on $S^2$; even in the zero 
instanton sector, after decompactification, no gap occurs for sufficiently
large values of ${\cal A}$. Only when ${\cal A}$ is small enough, the
complete and the zero-instanton solution exhibit similar behaviours.
We emphasize that this statement has to be proven, since the limits
$N\to \infty$ and $\AA \to 0$ could in principle not commute.
For small area, only the term with $m=0$ contributes to the summation
Eq.~\re{sigmasomML}, and therefore we deduce $\th_{cr}^{\tiny{ML}}=2 \sqrt\AA$.
This is precisely the leading term in the expansion of Eq.~\re{gap}, so
that the two values of $\th_{cr}$ coincides at small area.
Let us now show that also $\s(\AA,\th)$
Eq.~\re{sol} and $\s^{\tiny{ML}}(\AA,\th)$ have a similar behaviour for small 
${\cal A}$. As $\th<\th_{cr}$, by defining
$\th=\l\,\th_{cr}$,
we can write
\beq\nome{smallarea}
\s^{\tiny{ML}} = \frac1{\pi\sqrt\AA}\sqrt{1-\l^2}\,.
\eeq
Notice that the last factor in Eq.~\re{smallarea} is finite as $\AA\to
0$.
By comparing Eq.~\re{smallarea} and Eq.~\re{sol}, we expect that the
solutions $(\hat{x},\hat{y})$ of Eq.~\re{trasc} behave for small ${\cal A}$ as
$(-\half+\frac{a}{\sqrt\AA}, \frac{b}{\sqrt\AA})$, $a$ and $b$ being
finite numbers.
As a matter of fact, solving the trascendental equation Eq.~\re{trasc}
with this ansatz in mind, we find
\beeq\nome{smallsol}
 \hat{x}_{\pm}&\simeq& -\frac12 \pm \frac1{\sqrt{\AA}} \, \sqrt{1-\l^2} \no\\
 \hat{y}&\simeq& \frac\l{\sqrt{\AA}}\,.
\eeeq
Thus, we come to the remarkable result
\beq\nome{coin}
\sigma=\frac{1}{\pi}\lp\hat x_+ +\frac{1}{2}\rp\simeq
\frac1{\pi\sqrt\AA}\sqrt{1-\l^2}\,,
\eeq
which is just Eq.~\re{smallarea}.

Its precise meaning is elucidated if we remember that $\s({\cal
A},\th)$ has a distribution character in the limit ${\cal A}\to 0$.
After introducing the test function $\varphi(\th)$, and defining
\beq
\label {functional}
(\s,\varphi)\equiv\int_{-\th_{cr}}^{\th_{cr}} d\th \, \varphi(\th)\,
\s(\AA,\th), 
\eeq
we get
\beq
\label{dist}
(\s,\varphi)\simeq \frac{2}{\pi}\int_{-1}^{1} d\l \,
\sqrt{1-\l^2}\, \varphi(\l\th_{cr})\, \stackrel{{\cal A}\to 0}
{\longrightarrow}\varphi(0),
\eeq
which shows indeed that, when ${\cal A}\to 0$, $\s(\AA,\th)\to \delta(\th)$ 
in the topology of distributions, both for the exact and the WML case.

On the other hand, at large areas, the genuinely perturbative solution 
($i.e.$ the one coming from
the equal-time quantization and corresponding to the zero-instanton
sector), becomes less and less reliable.

To gain a better insight, we consider again the zero-instanton sector
on $S^2$, before decompactification. Eq.~\re{risulta} teaches us that this 
is easily achieved by replacing ${\cal A}$ with $\frac{(A-\AA)\AA}{A}$.
In the equatorial condition $A-\AA=\AA$, Eq.~\re{sigmasomML}
becomes
\begin{equation}
\label{sigmasfera}
\sigma^{\tiny{ML}}_{S^2}=\frac{2}{\pi \sqrt{A}}\sum_{m}
\sqrt{1-\frac{(2\pi m+\theta)^2}{A}},
\qquad\quad -\pi \leq \theta \leq \pi \,.
\end{equation}
It is amusing to notice that the critical value for $A$ is now again
$A_{cr}=\pi ^2$ and that, below such a critical value, the ML
spectral distribution $exactly$ coincides with the one in the
weak DK phase on the sphere\cite{kaza}.

Actually, from Eq.~\re{risulta}, we can derive  a stronger result;
we can indeed follow the evolution of the critical area of the loop
as a function of the size of the sphere also for the zero-instanton
sector.
As a matter of fact, Eq.~\re{sigmasomML} becomes
\beq
\label{critica}
\s^{\tiny{ML}}(\AA_{cr}^{\tiny{ML}},\pi)=
\frac{\sqrt A}{\pi\sqrt{\AA_{cr}^{\tiny{ML}}(A-\AA_{cr}^{\tiny{ML}})}}\,
\sqrt{1-\frac{\pi^2 A}{4\AA_{cr}^{\tiny{ML}}(A-\AA_{cr}^{\tiny{ML}})}},
\eeq 
leading to
\beq
\label{acritica}
\AA_{cr}^{\tiny{ML}}=\frac{A}{2}\lq 1-\sqrt{1-\frac{\pi^2}{A}}\rq.
\eeq
Starting from the value $\AA_{cr}^{\tiny{ML}}=\frac{\pi^2}{2}$, when $A=\pi^2$,
the same as the one of the exact solution, in the decompactification
limit $A\to \infty$ we recover the threshold
$\AA_{cr}^{\tiny{ML}}=\frac{\pi^2}{4}$,
lower than the one of the exact case ($\AA_{cr}=4$).

\section{Conclusions}
\noindent
The main concern of this paper was to gain a better insight in the dynamics 
of $YM_2$ by considering the class of Wilson loops winding $n$ times
around  a closed smooth contour, either on a compact manifold ($S^2$)
or on the plane. 
As already noticed by several authors,
thanks to the invariance under area-preserving diffeomorphisms,
these loops carry enough information to characterize the theory
in the large-$N$ limit.

Our deepest result is probably Eq.~\re{dual}. It shows a curious interplay 
between two integral numbers, $N$ characterizing the internal symmetry
group and $n$ representing the number of windings along the contour.
In a geometrical language, the latter controls the displacement of the 
connection on the base manifold, the former is related to displacements
along the group fiber. In two dimensions, the winding $n$ has
essentially a topological character, as it appears in the $n$-th power 
of the group element $U_{C}$; the Wilson loop can then be
interpreted as a generalized Clebsch-Gordan coefficient intertwining
different $U(N)$ representations. This is, we believe, the root of
Eq.~\re{dual} and, of course, it entails far-reaching consequences.
In particular it relates the abelian-like exponentiation for $U(N)$
when $n=1$ to the genuine triviality of the Maxwell theory ($N=1$)
for any $n$.

All these features are reproduced in a perturbative context. Actually 
concrete examples of different colour and geometrical interplays are 
explicitly exhibited, according to the different expressions used for 
the vector propagator (either WML or 't Hooft). When the 't Hooft propagator
is used the perturbative series leads to the exact result. However,
for $n>1$, the Wilson loop feels the non-abelian nature of the theory.
The winding number $n$ probes its colour content.
The related light-front vacuum, although simpler than the one in the
equal-time quantization, as it automatically takes instanton contributions
into account, cannot be considered trivial.

A more intriguing result concerns the limit $n \to \infty$ keeping
$n^2 {\cal A}$ fixed, namely the small-area limit. In such a limit
(Eq.~\re{granden}), one exactly recovers the contribution of the 
zero-instanton sector, namely Eq.~\re{risultatino}.
Intuitively this can be understood by remembering that instantons
possess a typical length scale measured by their size. 
One might then be
tempted to conclude that the phase transition at large $N$ is completely 
driven by instantons. Unfortunately this characterization is true only for 
small areas, when instantons cannot contribute. In Sect.~IV it is shown that, 
for areas larger than a critical threshold, the spectral function
which characterizes the eigenvalue distribution in the exact solution,
when computed retaining only the zero-instanton sector, does not
exhibit any gap. 
At large areas, the genuinely perturbative solution ($i.e.$ the one coming from
the equal-time quantization and corresponding to the zero-instanton
sector), becomes less and less reliable, as
expected on a general ground. It is then quite remarkable that the
perturbative light-front resummation succeeds in reproducing not only
the correct string tension, but also the non-trivial corrections
(see Eq.~\re{wilmorenice}) due to colour-winding factors, the winding 
number $n$ carrying the information
about the non-abelian topology.

One should perhaps conclude with a comment concerning higher dimensional
cases, in particular the dimension $d=4$.
We believe that most of our results are typical of the two-dimensional case.
Perturbation theory at $d=2+\epsilon$ is discontinuous in the limit
$\epsilon \to 0$ \cite{noi1,belli}; on the other hand the invariance under 
area-preserving diffeomorphisms is lost when $d>2$. In a perturbative
picture the presence of massless ``transverse'' degrees of freedom 
(the ``gluons'') forces a causal
behaviour upon the relevant Green functions, whereas in the soft (IR) limit
they get mixed with the
vacuum. The light-front vacuum, which also in two dimensions is far from
being
trivial, in higher dimensions 
is likely to be simpler only as far as topological degrees of
freedom are concerned. Of course there is no reason why it should coincide
with the physical vacuum since, after confinement, the spectrum
is likely to contain only massive excitations.  
Moreover, to be realistic, ``matter'' should be introduced, both in
the fundamental and in the adjoint representation. Therefore, before going to
higher dimensions, our two-dimensional considerations should perhaps be
generalized to the case in which ``matter'' is present.
Although many papers have appeared to this regard in the recent literature,
we feel that further work is still needed to reach
a complete understanding.

\section{Appendix A: \bom{\Wn} \bom{\OO (\gq)} with `t Hooft prescription}
In this appendix we explicitly compute $\Wn$ at order $\OO (g^4)$.  We
have seen in Sect.~III that at this order all diagrams come with the
prefactor given by Eq.~\re{prefactor}. We now have to group all possible
diagrams in different classes according to their contribution either
to $\tr \lq T^a\,T^a\,T^b\,T^b \rq$ (non-crossed diagrams) or to $\tr
\lq T^a\,T^b\,T^a\,T^b\rq$ (crossed diagrams), and to the area factor.
We then have to specify how many of them belong to each class
(depending on the number of windings $n$) and the area factor 
they produce~\footnote{Recall it is either $\half$ or 1 depending on
whether there are nested integrals in the loop variables or not.}.  
Such numbers are
summarized in the following tables: Table~\ref{noncros} refers to non-crossed
diagrams depicted in Fig.~2, whereas Table~\ref{cros} to crossed ones
depicted in Fig.~3.
Finally, summing up  the contributions from all loops, each multiplied by
the prefactor \re{prefactor}, the proper trace and area factor, and
taking into account the multiplicity, we recover the result announced
in \re{sum4} for the Wilson loop with $n$ windings at order $\OO (g^4)$.

\vskip 1.0truecm

\section{Appendix B}
In this appendix we will sketchily show that Eq.~\re{trasc} admits
two solutions  and, of these, only one  can be  encircled by the integration
contour.
Let us start from Eq.~\re{trasc}, which, after inserting $t=x+iy$,
reads
\beq\nome{imre}
\frac{1+x+iy}{x+iy} \, \ed{-\AA\,(x+\frac12)}\, \ed{i\lp\th -\AA y\rp} =1\,.
\eeq
The first obvious symmetry to be noticed is $(x,y,\theta)\to (x,-y,-\theta)$. 
In addition, by taking the absolute value
of Eq.~\re{imre},  the symmetry $x\to -x-1$ is also manifest.
Combining Eq.~\re{imre} and its complex conjugate, we obtain an equation
for the shape of
the boundary which separates the region of convergence of
Eq.~\re{rappres} and the forbidden region
\beq\nome{bound}
y^2=-( x^2 + x + \half ) + ( x + \half )\coth{(\AA ( x +
\half))} \,, \qquad {\rm for} \quad x\neq -\half\,.
\eeq
It follows
\beq\nome{dery}
\fracd{dy^2}{dx}= -\frac{x + \half}{\sinh^2{(\AA( x + \half ))}}\, f(x;\AA)\,,
\eeq
where
\beq\nome{f}
f(x;\AA)= \AA -\frac{\sinh{( \AA( 2x+1 ))}}{2x+1} + 
2 \sinh^2{(\AA (x + \half ))}\,.
\eeq
It is easy to check that the same function $f$ appears in the derivatives
of the coordinates $x$ and $y$ with respect to $\th$. From Eq.~\re{imre}
we infer
\beeq\nome{parder}
\fracd{\p x}{\p \th} &=&\fracd{\p y}{\p \th} \,
\fracd{y\, g(x,y)}{f(x;\AA)}\,, \no \\
\fracd{\p y}{\p \th} &=&\fracd{f(x;\AA)}{f^2(x;\AA) + y^2 g^2 (x,y)}\,,
\eeeq
with
$$
g(x,y)=\frac1{(x+1)^2+y^2}-\frac1{x^2+y^2}\,.
$$
By comparing Eqs.~\re{dery}, \re{parder} we deduce
\beeq\nome{sign}
{\rm sign}\fracd{\p y}{\p \th} &=&- {\rm sign}\fracd{dy^2}{dx} \qquad \,
{\rm if}\quad  x>-\half\,, \no \\
{\rm sign}\fracd{\p y}{\p \th} &=& {\rm sign}\fracd{dy^2}{dx} \qquad \quad
{\rm if} \quad x<-\half\,.
\eeeq
We conclude that, if for $\th=\th_0$ a root of Eq.~\re{trasc} exists, 
for $\th>\th_0$ its
ordinate $y_r(\th,\AA)$ follows the boundary continuously.
Moreover, Eq.~\re{imre} tells us that for $\th=0$ 
there are two roots
sitting on the $x$-axis ($y_r=0$), whose $x$ coordinates are the solutions of
the following equation
\beq\nome{thzero}
\frac{x+1}x = \ed{\AA\,\lp x+\half \rp}.
\eeq
One can check that Eq.~\re{thzero} admits only two solutions, for any value
of $\AA$, symmetric with respect to the axis $x=-\half$.
Clearly, when $x$ approaches the value $-\half$, the condition $y^2
\geq 0$ in Eq.~\re{bound} can be fulfilled only if $\AA \leq 4$.
Therefore, we start at  $\th=0$ with two roots on the real axis which 
move continuously along   the boundary   as $\th$ increases if
$\AA > 4$, whereas for $\AA \leq 4$ there is a critical value of
$\th$, let us call it $\th_{cr}$, for which the two roots reach the
axis $x=-\half$ and then remain on it, moving in the imaginary direction.
Hence a gap is originated, since
$$
\s (\AA,\th)=0 \quad\quad {\rm for} \quad \th\geq\th_{cr}\,, \quad 
                                          \AA \leq 4\,.
$$
At this stage we present an hand-waving argument to show that only one
of the two roots is encircled by the integration contour. Nevertheless
a rigorous proof based on a thorough examination of Eq.~\re{ineq} can
be given.
Although the region of  convergence of Eq.~\re{rappres} appreciably
varies  according to the value  of $\AA$, it is always made up of
disconnected pieces
and the axis $x=-\half$ represents  a border line  between a domain of
convergence and a forbidden region. As a consequence, the integration
contour, which encircles the origin of the complex plane (lying in the
forbidden region), is forced
not  to cross such an axis and to capture just one of the two roots
(remember they are symmetric with respect to the exchange $x\to -x-1$).
This root in turn
and the origin become the two branch points
of the logarithm in Eq.~\re{single}.

Let us now determine the exact value of $\th_{cr}$ in case $\AA<4$.
The criticality is reached when $\hat{x}=-\half$, so that we evince
from Eq.~\re{bound} 
$$
y_{cr}=\pm \sqrt{\frac1\AA - \frac14}\,.
$$
Then Eq.~\re{imre} becomes
\beq \nome{sistema}
\lgr 
\begin{array}{ll}
-\half & =\half \cos{(\th_{cr} -A y_{cr})} - y_{cr} \sin{(\th_{cr} -A
y_{cr})} \\
y_{cr} &= \half \sin{(\th_{cr} -A y_{cr})} + y_{cr} \cos{(\th_{cr} -A
y_{cr})}
\end{array}
\right.
\eeq
Eventually, by inserting the value of $y_{cr}$, the solution is
straightforwardly found
\begin{equation}
\theta_{cr}=\sqrt{{\cal A}-\frac{{\cal A}^2}{4}}+\arccos\lp 
1-\frac{{\cal A}}{2}\rp\,.
\end{equation}

\vskip 1.0truecm

${\bf Acknowledgement}$
Discussions at an early stage of this work with G. Nardelli
are acknowledged. One of us (A.B.) wishes to thank Y. Frishman
and G. Miller for their hospitality at the
Physics Departments of the Weizmann Institute (Rehovot) and
of the University of Washington (Seattle), respectively,
while part of this work was done.

\begin{table}[h]
 \begin{center}
  \begin{tabular}{|l|c|c|}
Diagrams & number & area factor\\ \hline
Fig.~1a & $n$ & $\half$ \\ \hline
Fig.~1b & $2n (n-1)$ & $\half$ \\ \hline
Fig.~1c & $n (n-1)$ & $\half$ \\ \hline
Fig.~1d & $\frac{n(n-1)}2$ & 1 \\ \hline
Fig.~1e & $\frac{n(n-1)}2$ & 1 \\ \hline 
Fig.~1f & $n (n-1) (n-2)$ & $\half$ \\ \hline
Fig.~1g & $n(n-1)(n-2)$ & 1 \\ \hline
Fig.~1h & $\frac{n(n-1)(n-2)}2$ & 1 \\ \hline
Fig.~1i & $\frac{n(n-1)(n-2)(n-3)}3$ & 1
   \end{tabular}
  \caption{\label{noncros}Classes of non-crossed diagrams contributing
  to $\Wn$ $\OO (g^4)$ with counting and area factor.}
 \end{center}
\end{table}

\begin{table}[h]
 \begin{center}
  \begin{tabular}{|l|c|c|} 
Diagrams & number & area factor\\ \hline
Fig.~2a & $2n (n-1)$ & $\half$ \\ \hline
Fig.~2b & $n (n-1) (n-2)$ & $\half$ \\ \hline
Fig.~2c & $\frac{n(n-1)(n-2)}2$ & 1 \\ \hline
Fig.~2d & $\frac{n(n-1)(n-2)(n-3)}6$ & 1 
   \end{tabular}
  \caption{\label{cros}Classes of crossed diagrams contributing to
  $\Wn$ $\OO (g^4)$ with counting and area factor.}
 \end{center}
\end{table}

\begin{center}
\begin{figure}
\epsfysize=10cm
\epsfbox{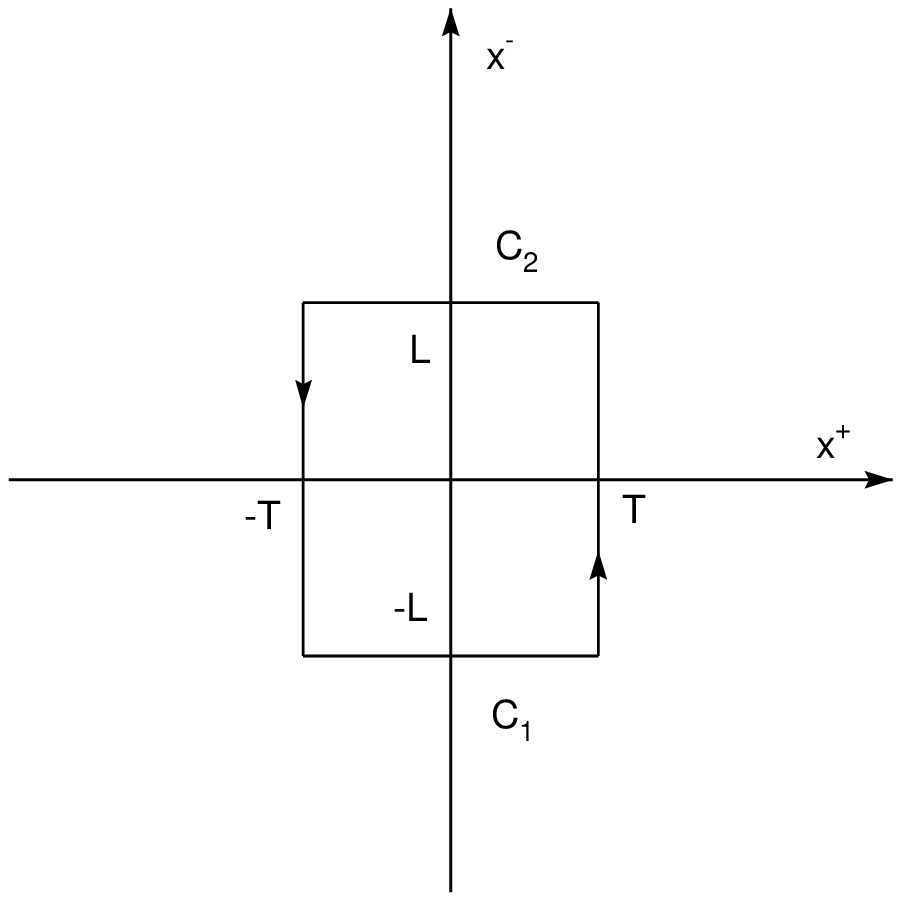}
\caption{Contour.}
\end{figure}
\end{center}
\begin{center}
\begin{figure}[htbp]
\epsfysize=4.5cm
\epsfbox{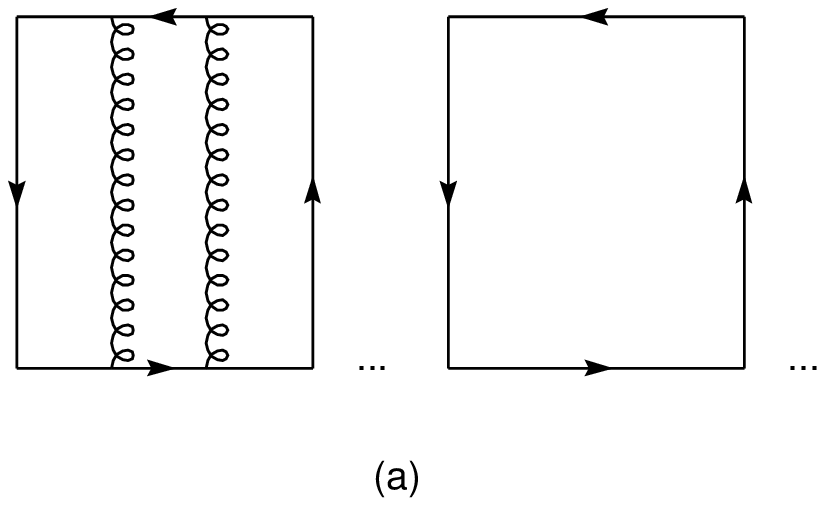}
\end{figure}
\begin{figure}[htbp]
\epsfysize=4.5cm
\epsfbox{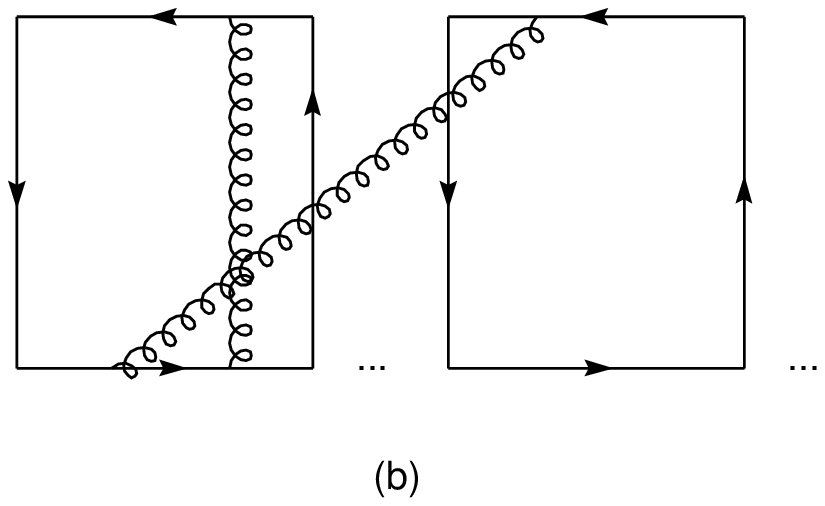}
\end{figure}
\begin{figure}[htbp]
\epsfysize=4.5cm
\epsfbox{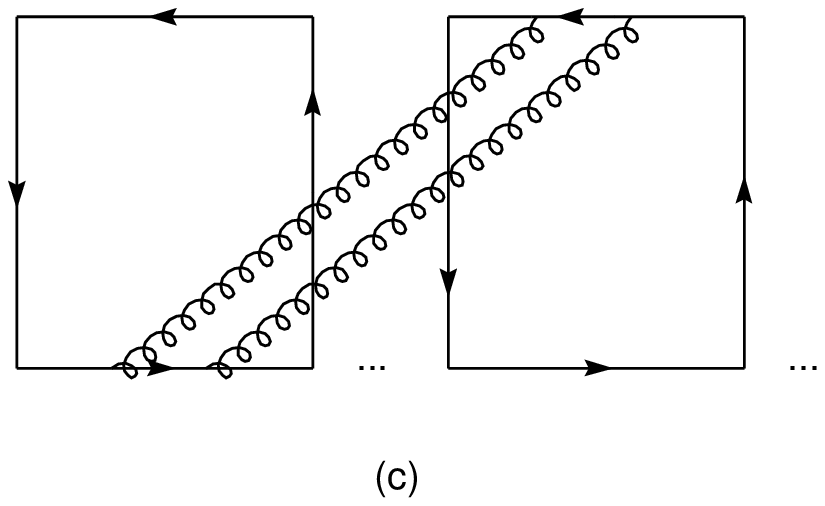}
\end{figure}
\begin{figure}[htbp]
\epsfysize=4.5cm
\epsfbox{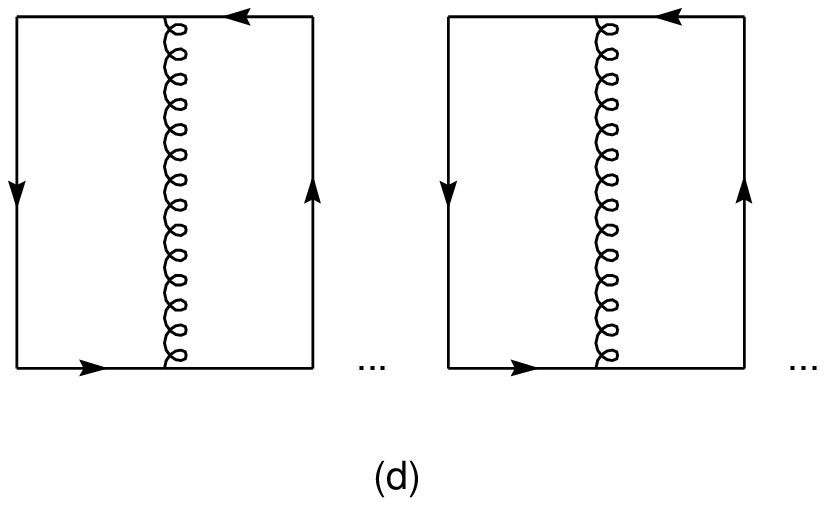}
\end{figure}
\begin{figure}[htbp]
\epsfysize=3.5cm
\epsfbox{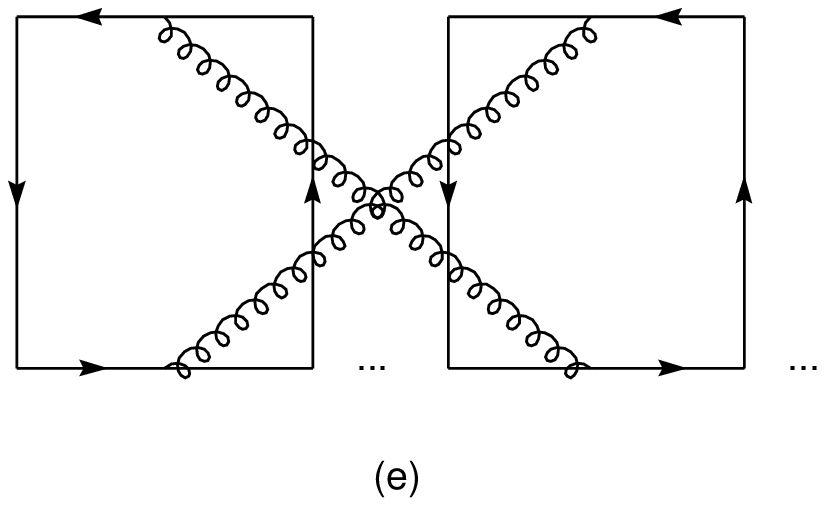}
\end{figure}
\begin{figure}[htbp]
\epsfysize=3.5cm
\epsfbox{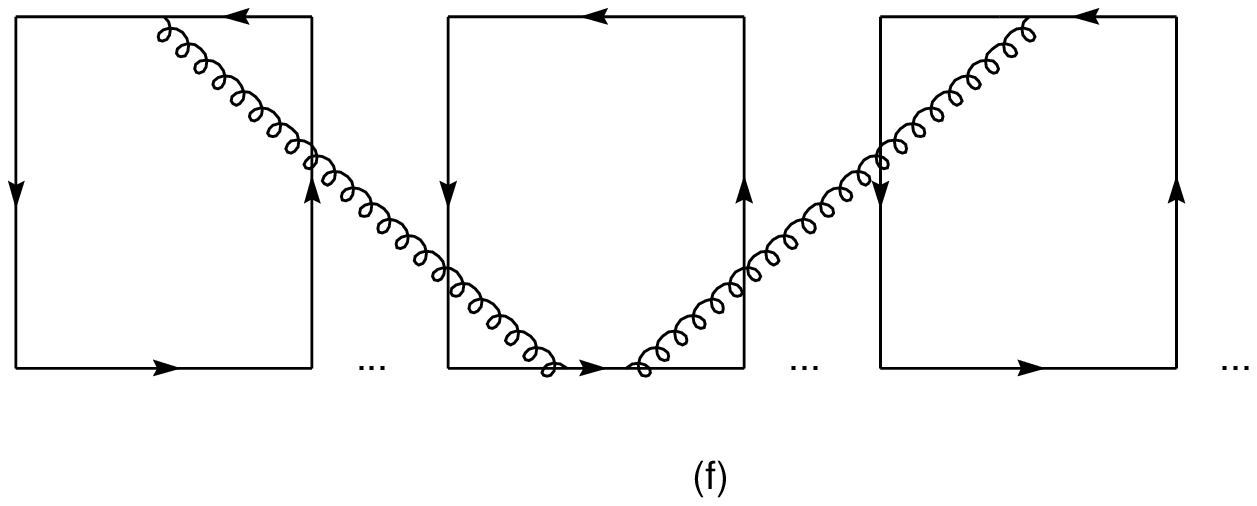}
\end{figure}
\begin{figure}[htbp]
\epsfysize=3.5cm
\epsfbox{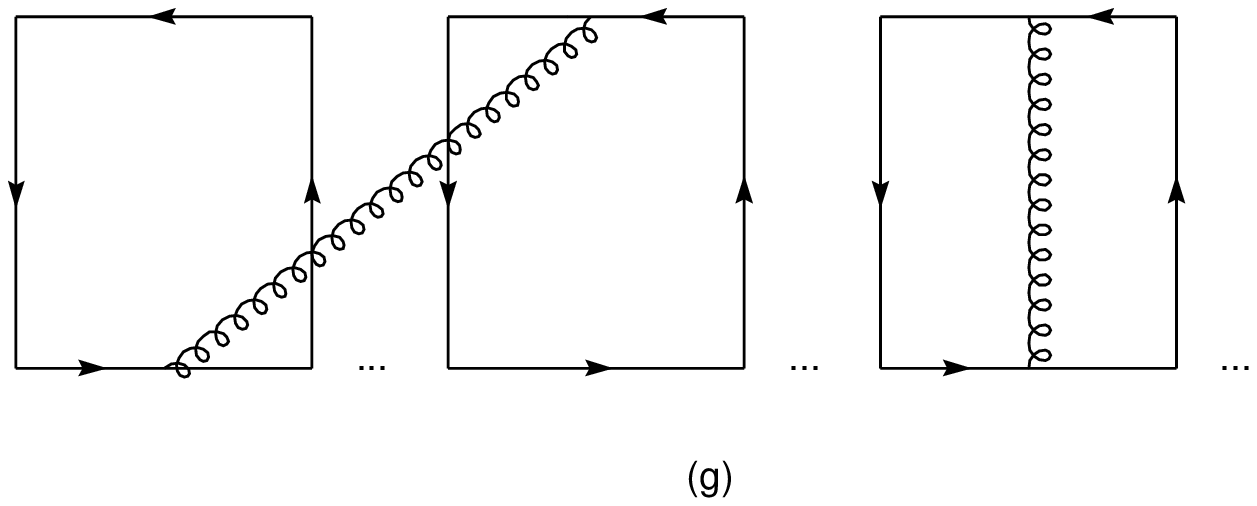}
\end{figure}
\begin{figure}[htbp]
\epsfysize=3.5cm
\epsfbox{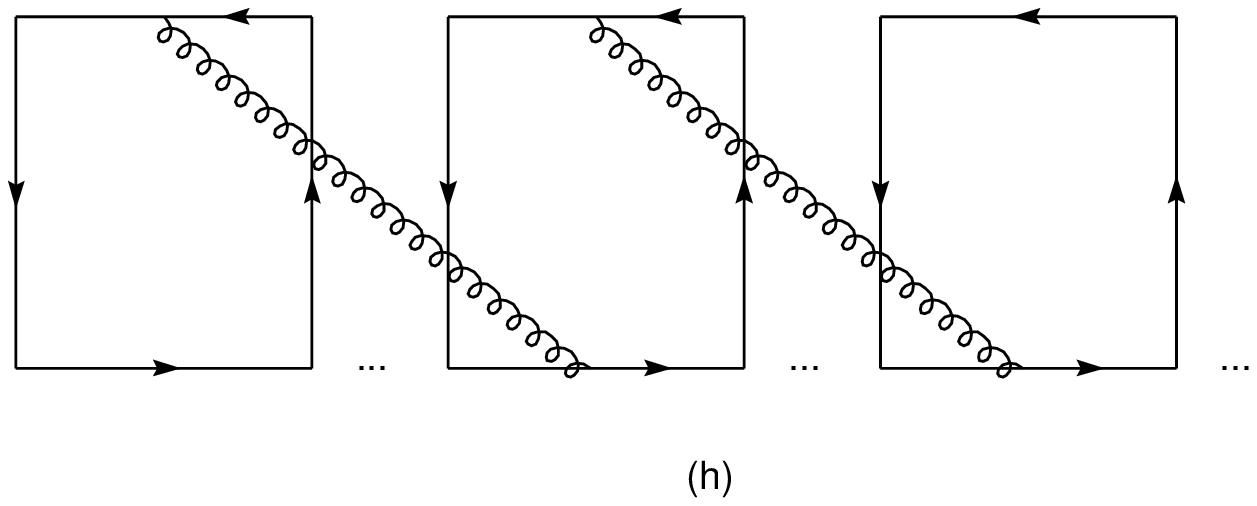}
\end{figure}
\begin{figure}[htbp]
\epsfysize=4cm
\epsfbox{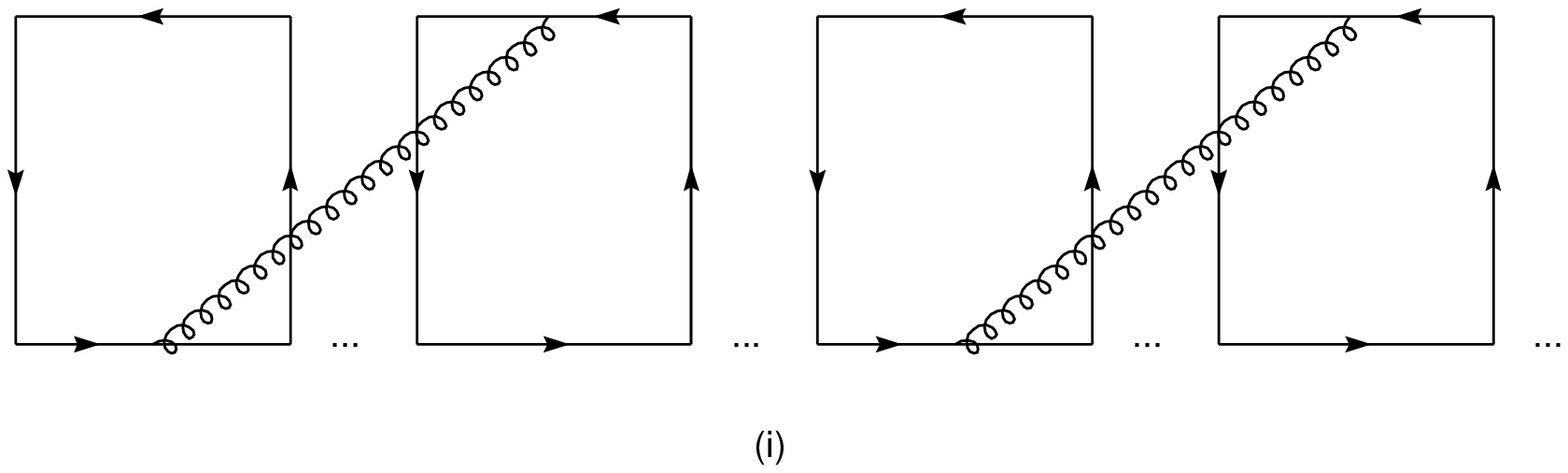}
\caption{Non-crossed graphs. The $i^{\rm th}$ sheet
corresponds to the $i^{\rm th}$ winding.}
\end{figure}
\end{center}

\begin{center}
\begin{figure}[htbp]
\epsfysize=4cm
\epsfbox{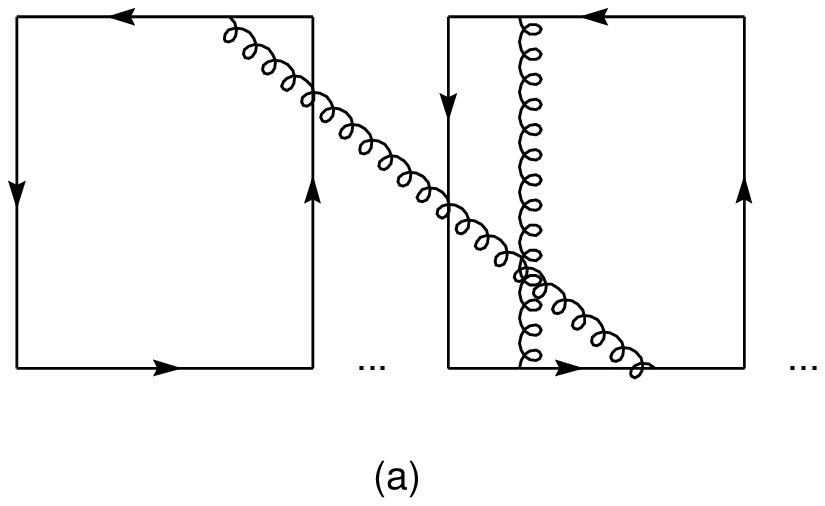}
\end{figure}
\begin{figure}[htbp]
\epsfysize=4cm
\epsfbox{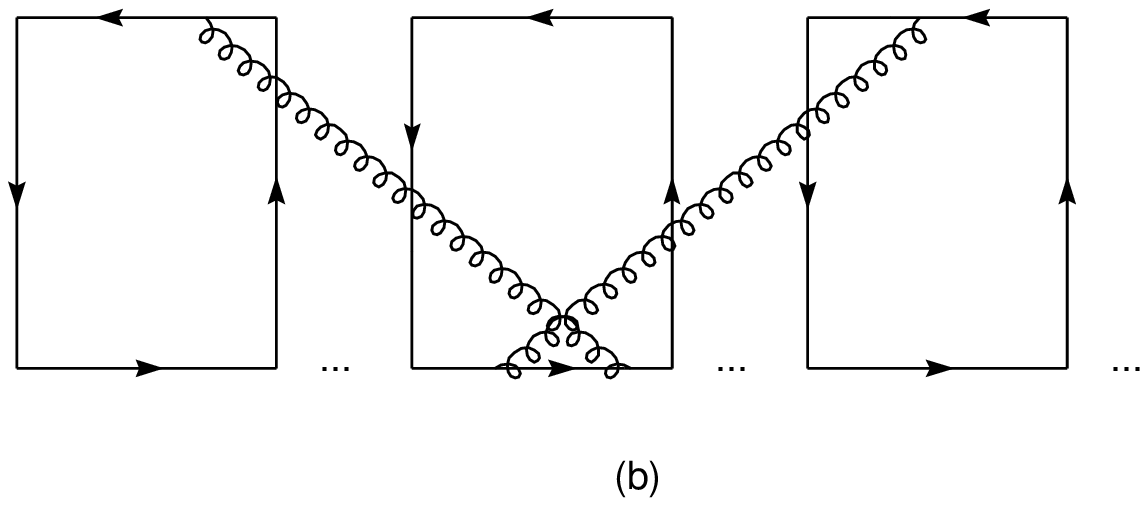}
\end{figure}
\begin{figure}[htbp]
\epsfysize=4cm
\epsfbox{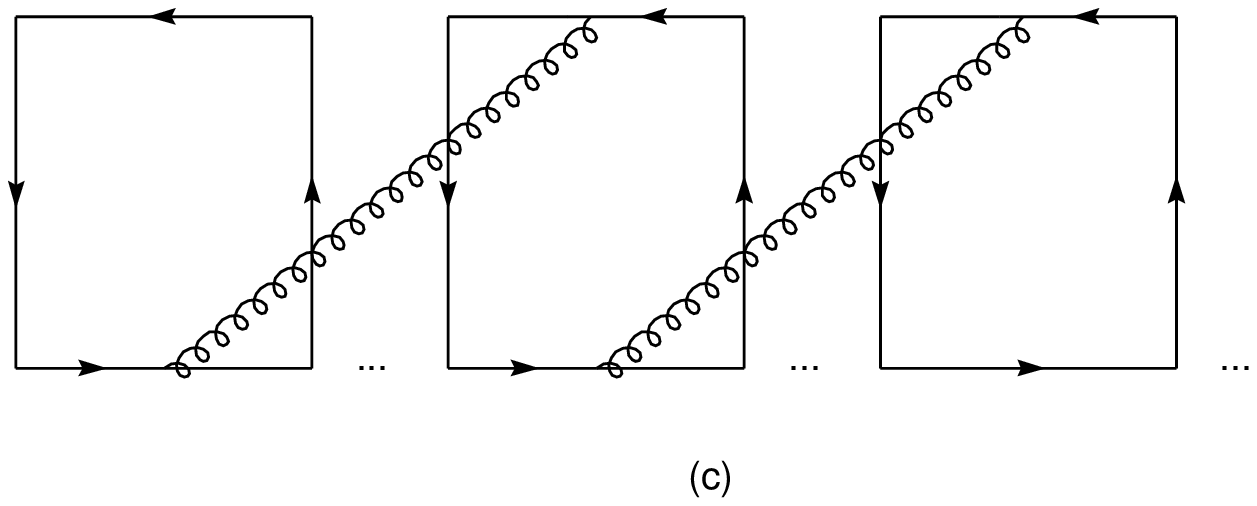}
\end{figure}
\begin{figure}[htbp]
\epsfysize=4.5cm
\epsfbox{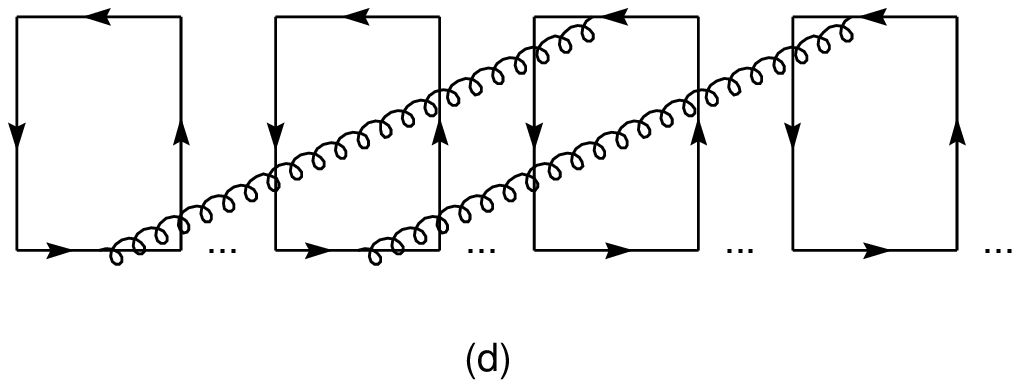}
\caption{Crossed graphs.}
\end{figure}
\end{center}

\vfill\eject

\end{document}